# Theory of Two-Photon Absorption with Broadband Squeezed Vacuum

*Michael G. Raymer*[1,2] * *and Tiemo Landes*[1,2]

[1] *Department of Physics, University of Oregon, Eugene, OR 97403, USA*
[2] *Oregon Center for Optical, Molecular and Quantum Science, University of Oregon, Eugene, OR 97403, USA*
* *raymer@uoregon.edu*

**Abstract:** We present an analytical quantum theoretic model for non-resonant molecular two-photon absorption (TPA) of broadband, spectrally multi-mode squeezed vacuum, including low-gain (isolated entangled photon pairs or EPP) and high-gain (bright squeezed vacuum or BSV) regimes. The results are relevant to the potential use of entangled-light TPA as a spectroscopic and imaging method. We treat the scenario that the exciting light is spatially single-mode and is non-resonant with all intermediate molecular states. In the case of high gain, we find that in the case that the linewidth of the final molecular state is much narrower than the bandwidth of the exciting light, bright squeezed vacuum is found to be equally (but no more) effective in driving TPA as is a quasi-monochromatic coherent-state (classical) pulse of the same temporal shape, duration and mean photon number. Therefore, in this case the sought-for advantage of observing TPA at extremely low optical flux is not provided by broadband bright squeezed vacuum. In the opposite case that the final-state linewidth is much broader than the bandwidth of the BSV exciting light, we show that the TPA rate is proportional to the second-order intensity autocorrelation function at zero time delay $g^{(2)}(0)$, as expected. We derive and evaluate formulas describing the transition between these two limiting cases, that is, including the regime where the molecular linewidth and optical bandwidth are comparable, as is often the case in experimental studies. We also show that for $g^{(2)}(0)$ to reach the idealized form $g^{(2)}(0) = 3 + 1/\bar{n}$, with $\bar{n}$ being the mean number of photons per temporal mode, it is required to compensate the dispersion inherent in the nonlinear-optical crystal used to generate the BSV.

## 1. Introduction

Two-photon absorption (TPA) is a widely method in spectroscopy, as it can yield results that are not available via linear (one-photon) absorption methods. For 'classical' (coherent state) fields and broadband absorbers, the rate of TPA is proportional to the square of the instantaneous intensity of the electromagnetic field. To achieve high intensities while keeping the average flux low, short pulses can be used to increase the efficiency of the process. However, for ultrashort pulses and narrowband absorbers, the increase in efficiency is limited by the spectral overlap between the two-photon transition and the driving field. To increase the TPA efficiency, it has been proposed to use broad-band photon pairs that are time-frequency quantum entangled—such that the sum of their frequencies is equal to the material's two-photon resonance frequency—to drive two-photon absorption. Recently, efforts to implement two-photon spectroscopy and imaging have met with some skepticism regarding its practicality, due to the extremely low events rates predicted by standard theories and supported by some recent experiments. [[1], [2], [3]]

This paper addresses whether significant advantages can be obtained in this regard by instead using so-called bright squeezed vacuum (BSV) to drive TPA.

The concept of entangled two-photon absorption (ETPA) is well established theoretically in the low-flux regime of isolated photon pairs (in which distinct pairs do not overlap within the field's correlation time or within the molecule's response time). [4, 5, 6, 7, 8, 9] Fei et. al. emphasized a simple heuristic model in which the rate of ETPA is represented approximately as [6]

$$R = \sigma_e \frac{F}{A_0} + \sigma^{(2)} \left(\frac{F}{A_0}\right)^2, \qquad (1)$$

where $F$ is the total photon rate [photons s$^{-1}$], $A_0$ is the effective beam area [m$^2$], $\sigma_e$ is called the ETPA cross section [m$^2$] arising from isolated-entangled-pair photons, and $\sigma^{(2)}$ is the TPA cross section [m$^4$ s] arising from accidental coincidences of photons. In the case of monochromatic 'classical' light, $\sigma^{(2)}$ is equal to the conventional TPA cross section, first derived by Maria Göppert-Mayer. [10] For molecules in solution $\sigma^{(2)}$ is typically exceedingly small—on the order of 1 to 1,000 GM (where 1 GM = 10$^{-58}$ m$^4$s). [11]

In a previous study we derived an upper bound on the (low-flux) isolated-entangled-pair cross section $\sigma_e$ in the case that the molecular final-state TPA linewidth is narrow compared to the entangled photon pair's bandwidth. Using perturbation theory under standard assumptions assuming homogeneously broadened molecular energy eigenstates yields the bound [1, 12]

$$\sigma_e \leq \frac{\sigma^{(2)}}{A_0} 2B, \qquad (2)$$

where $B$ is the full bandwidth of the EPP spectrum in units of Hz. Quantitative estimates using this upper bound indicate that with EPP fluxes limited to the isolated-pair regime and realistic sample concentrations event rates are orders of magnitude below the detection threshold of typical photon-counting systems. This prediction is consistent with experimental efforts reported in [2] and [3], although other experiments have seemed to indicate otherwise. [13, 14]

While it is the case that temporal-spectral correlation embodied in quantum entanglement can significantly increase the TPA rate as a result of effective spectral compression at the two-photon resonance frequency, the above analysis predicts that the rates of ETPA for typical molecules are nevertheless too small for practical experimental observation in the low-flux isolated-pair regime. [12, 15, 16]

To overcome the too-small TPA rates when using entangled photon pairs in the low-flux isolated-pair regime, it is natural to wonder if using bright squeezed vacuum (BSV) states of light can yield large enhancements while creating more-readily observable ETPA rates. BSV is defined as a squeezed state of light that has zero mean field but a high number of photon pairs per mode. [17] Prior theories that address ETPA in the high-gain PDC (that is, BSV) regime include those of Dayan [18] and Schlawin and Mukamel [7]. Dayan derived expressions for two-photon interactions (TPA as well as sum-frequency generation) induced by broadband down-converted light that was generated from a narrow-band (long-pulse) pump laser. Dayan concluded that for such time-frequency-entangled sources the rate of such processes is the sum of two contributions – a 'coherent' term, which depends on the coherent overlap of temporal-spectral components of the field and thus is affected strongly by dispersion-induce time delays; and an 'incoherent' term, which depends only on the overlap of the field's intensity with itself and is thus less sensitive to dispersion. The characteristic time scale of the coherent term is the inverse bandwidth of the EPP field, while the characteristic time scale of the incoherent term is the much longer duration of the intensity envelope. Furthermore, Dayan pointed out that at low flux, the rate of TPA driven by the coherent term scales linearly with the mean photon flux, while the incoherent term scales quadratically, making a connection with the previously cited study by Fei et. al. in Eq.(1).

Schlawin and Mukamel considered ETPA in a different regime, where the PDC producing broadband BSV is pumped by an ultrashort (ps or fs) laser pulse. [7] An accurate description in this case requires the use of a singular-value decomposition to discover the appropriate time-frequency (temporal) mode basis in which to represent the set of independent two-mode squeezing (Bogoliubov) transformations present in the BSV field that drive the TPA. They found similar scaling of ETPA rates with EPP flux as found by Dayan and expressed in Eq.(1).

The present paper presents a reexamination and clarification of the ETPA problem along lines most similar to Dayan's, with several key differences: We use a single-spatial-mode model for multi-temporal-mode squeezed light that enables the TPA calculations to be carried out explicitly and analytically while making only well-controlled approximations. Our results confirm that the heuristic formula in Eq.(1) is accurate under appropriate conditions, as our rigorous result Eq.(44) reproduces it along with explicit expressions for the proportionality factors. We derive new formulas describing how the quantum enhancement of ETPA arises in both the low- and high-flux regimes of PDC; and we derive a simple expression for the cross-over between linear and quadratic scaling with flux, consistent with previous results, for example [5]. We clarify the roles of the coherent and incoherent contributions to the TPA rate under conditions of broad or narrow TPA final-state linewidths, and how these relate to the conventional understanding based on the value of $g^{(2)}(0)$.

The above-discussed theories focus on the scenario that the exciting light is non-resonant with all intermediate molecular states. That is also our focus here. The case of resonant intermediate states was treated recently by Drago and Sipe [19], whose results reduce to many of ours in the non-resonant limit.

We confirm the known result that when the molecular final-state linewidth is very broad, such that the molecule responds instantaneously to fluctuations of the light, the TPA rate is proportional in general to the second-order (intensity) correlation function $g^{(2)}(0)$. For spectrally multimode BSV we prove that compensation of the dispersion inherent in the nonlinear-optical crystal used to generate the BSV is required to reach a three-fold rate enhancement ($g^{(2)}(0) = 3$) relative to coherent-state light. A three-fold enhancement has been predicted or measured previously for BSV relative to coherent-state or pseudo-coherent-state light. [20, 21, 22, 23, 24] We determine the amount of dispersion compensation needed to reach this optimum using a realistic model of spectrally multimode BSV.

We address the question: In the opposite limit, when the molecular TPA linewidth is narrow compared to a much broader bright squeezed vacuum spectrum, how great an enhancement of TPA does BSV provide relative to a coherent-state pulse? We derive and evaluate formulas describing the transition between the narrowband and broadband limits, including the regime where the molecular linewidth and optical bandwidth are comparable, as is often the case in experimental studies of entanglement-enhanced molecular TPA.

We find that in this limit the BSV is equally effective in driving TPA as a quasi-monochromatic coherent-state (classical) pulse of the same temporal shape and mean photon number. Thus, in this case there is no relative enhancement. Therefore, the sought-for advantage of observing TPA at low optical flux continues to be elusive.

## 2. Model for entangled two-photon absorption

We first review briefly the 'standard' model for two-photon absorption, used in many successful studies. While the historically first treatment of TPA used second-order perturbation theory for molecular pure-state quantum amplitudes and a final density of states [10] as reviewed in [11], we follow the method in which fourth-order perturbation theory is applied to the molecular-state density operator, allowing for a treatment of homogeneous dephasing linewidths of the transitions involved. [7, 8, 25, 26]

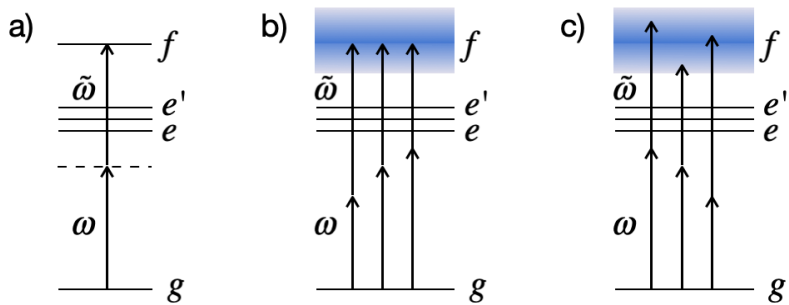

Fig. 1 (a) 'Direct' two-photon-excitation process in which photons of frequency $\omega$ and $\tilde{\omega}$ are absorbed simultaneously with no resonant intermediate state. (b) Coherent contribution of BSV which arises from frequency-anti-correlated photon pairs. (c) Incoherent contribution of the BSV, which arises from frequency-non-correlated pairs, contributes significantly to TPA only if the molecular line (shaded region) is broad enough to respond to non-optimal frequency combinations.

As reviewed in [15] used in [1, 12], when the sum of the two photon's frequencies is near resonance with the TPA transition, and the individual fields are far from any intermediate resonance, as in **Fig. 1**, the dominant term in the perturbation expansion is the so-called double-quantum coherence (DQC) term, which represents direct excitation to the $f$ state by simultaneous absorption of two photons without creating 'real' population in the intermediate states. We consider this term only, which yields for the probability to find the molecule in the *f* state following the excitation pulse

$$P^{TPA} = \sigma^{(2)} \frac{\gamma_{fg}}{A_0^2} \mathrm{Re} \int d\omega' \int d\omega \int d\tilde{\omega} \frac{C^{(4)}(\omega',\tilde{\omega}',\omega,\tilde{\omega})}{\gamma_{fg} - i\omega_{fg} + i\omega + i\tilde{\omega}}, \tag{3}$$

where $C^{(4)}$ is given below. For compactness we denote $\tilde{\omega}' = \omega + \tilde{\omega} - \omega'$ and denote $d\omega = d/2\pi$. Here $\omega_{fg}$ is the frequency and $\gamma_{fg}$ the dephasing rate between the ground state $g$ and final state *f*. The variable frequencies of the two photons that lead to excitation are $\omega, \tilde{\omega}$. The effective beam area at the molecule's location is $A_0$. Note again that we treat the case of spectrally multimode but spatially single-mode squeezed vacuum.

The conventional TPA cross section is [11]

$$\sigma^{(2)} = \left(\frac{\omega_0}{\hbar\varepsilon_0 nc}\right)^2 \frac{1}{2\gamma_{fg}} \left|\sum_e \frac{d_{ef} d_{ge}}{\omega_{eg} - \omega_0}\right|^2 \tag{4}$$

for excitation by monochromatic light far from resonance with all intermediate molecular states, where $\omega_0$ is the central frequency (rad/s) of the exciting field's spectrum, the electric-dipole matrix elements are $d_{jk}$, $n$ is the medium's refractive index at the center frequency, $\varepsilon_0$ is the vacuum permittivity, and $c$ is the vacuum speed of light. Equation (3) is consistent with that, for example, in [7] in the case of non-resonant intermediate states, as we treat here.

The nature of the exciting field is embodied in the four-frequency correlation function

$$\begin{aligned}C^{(4)}(\omega',\tilde{\omega}',\omega,\tilde{\omega}) &= Tr\left[\hat{\rho}_F \hat{c}^\dagger(\omega')\hat{c}^\dagger(\tilde{\omega}')\hat{c}(\omega)\hat{c}(\tilde{\omega})\right] \\ &\to \langle\Psi|\hat{c}^\dagger(\omega')\hat{c}^\dagger(\tilde{\omega}')\hat{c}(\omega)\hat{c}(\tilde{\omega})|\Psi\rangle\end{aligned}, \tag{5}$$

where $\hat{\rho}_F$ is the density operator for the field state, and the last line applies in the case of a pure state $\Psi$. As in **Appendix A**, the creation and annihilation operators $\hat{c}^\dagger, \hat{c}$ are related to the electric field operator at the location of the molecule, to good approximation, by

$$\hat{E}^{(+)}(t) = L_0 \int d\omega \, \hat{c}(\omega) e^{-i\omega t}, \tag{6}$$

where $L_0 = (\hbar\omega_0 / 2\varepsilon_0 nc A_0)^{1/2}$. For a single molecule located at $\mathbf{r}_0$, the proper definition of the effective beam area is given by $(1/A_0)^{1/2} \equiv u(\mathbf{r}_0)$, where the mode amplitude is normalized in the transverse spatial coordinates, $\int |u(\mathbf{r})|^2 d^2x = 1$. We assume the field is polarized linearly with a single fixed orientation, and the $d_{jk}$ in Eq.(4) represent the electric-dipole matrix elements projected onto the field polarization. The bosonic commutator is $[\hat{c}(\omega), \hat{c}^\dagger(\omega')] = 2\pi\delta(\omega - \omega')$.

## 3. Low- and high-gain squeezing regimes of PDC

In Type-0 or Type-I spontaneous parametric down conversion (PDC) in a second-order nonlinear optical crystal photon pairs are generated having the same linear polarization, leading to a large bandwidth determined by second-order dispersion. [27] These processes can be designed via phase matching to occur predominantly into a single collinear forward-traveling spatial mode, such that the photons have no distinguishing labels other than frequency. [28] Alternatively, phase matching for Type-0 or Type-I can be adjusted so co-polarized photon pairs are emitted and detected off axis on opposite sides of the pump beam; then a distinguishing label is the direction of propagation. Finally, Type-II PDC creates orthogonally polarized photon pairs either on or off axis.

We focus on the case of Type-0 or Type-I co-polarized, co-propagating photon pairs in the main part of the paper because of its simplicity. The other cases have similar predictions and are discussed in **Appendix B**.

To treat BSV-induced TPA that includes both low- and high-gain regimes, the Heisenberg picture is most useful to describe the action of the PDC crystal. As mentioned, an exact treatment requires numerical solutions of the propagation equations and is most generally described in terms of a singular-value decomposition (SVD) using temporal modes. [29, 30] Schlawin and Mukamel used SVD in a gaussian approximation to calculate frequency-resolved photon correlations and TPA probabilities. [7] Instead of SVD, Dayan used a quasi-steady-state approximation, valid if the pump laser pulse has an arbitrary shape but is slowly varying and long compared to the coherence time (inverse bandwidth) of the PDC light. [18]

We use a different approach, which allows deriving explicit closed-form expressions for the ETPA rate, while requiring only a few well-understood approximations. We consider that the pump for the PDC is continuous-wave (CW), allowing a straightforward solution of the broadband squeezing equations of motion. To model the PDC light as a pulse of duration $T$, we send the CW PDC beam through a shutter that opens suddenly and closes after a time $T$. as shown in **Fig. 2**. When assuming that the shutter opening time is long compared to the coherence time, this approach enables us to use Eq.(3) to calculate analytically the probability of excitation after a time $T$. The results are valid in both low- and high-gain regimes, and importantly allow us to understand quantitatively how the two regimes merge at moderate gain, leading to a derivation of the linear-and-quadratic scaling relation in Eq.(1). The same model, but without the time shutter, was used by Boitier, et. al., to describe two-photon interferometry with a two-photon-absorbing detector, and their results are consistent with ours in the case of an absorber linewidth much greater than the BSV bandwidth. [31] The figure also shows an optional dispersion-compensating device that may be inserted prior to interacting with the molecular sample.

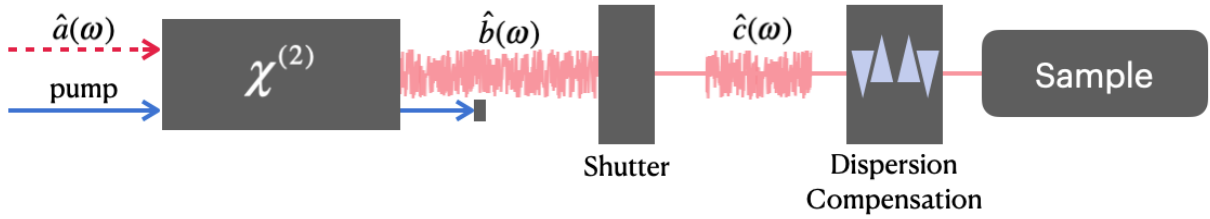

Fig. 2 Modelled experimental setup. The initial field (in the vacuum state) enters the second-order nonlinear optical crystal pumped by a CW laser and phase-matched for degenerate Type-0 or Type-1 spontaneous parametric downconversion (PDC), in which the colinear generated light has a single, linear polarization. The pump is blocked and the PDC passes through an OPEN-CLOSE shutter that is open for a duration $T$, then passes through an optional dispersion-compensating optic, and into the two-photon-absorbing molecular sample.

Referring to **Fig. 2**, the PDC crystal, with nonlinear coefficient $\chi^{(2)}(\omega)$, pumped by a monochromatic CW laser with field amplitude $E_{p0}$ and angular frequency $\omega_p = 2\omega_0$, causes a Heisenberg-picture transformation of the input (vacuum) field operators $\hat{a}(\omega)$ to output field operators $\hat{b}(\omega)$, then the field passes through a temporal shutter transforming the field operators to $\hat{c}(\omega)$. As reviewed in **Appendix A**, the squeezing transformation is given by the frequency-dependent two-mode squeezing transformation [32]

$$\hat{b}(\omega) = f(\omega)\hat{a}(\omega) + g(\omega)\hat{a}^\dagger(2\omega_0 - \omega) \ . \tag{7}$$

The gain functions are, for collinear Type-0 or Type-I phase matching and crystal length $z$,

$$\begin{aligned} f(\omega) &= \cosh[s(\omega)z] - i\frac{\Delta k(\omega)}{2s(\omega)}\sinh[s(\omega)z] \\ g(\omega) &= i\frac{\gamma(\omega)}{s(\omega)}\sinh[s(\omega)z] \end{aligned}, \tag{8}$$

where $\Delta k(\omega)$ is the phase mismatch of wavenumbers $k(\omega)$

$$\Delta k(\omega) = k_p - 2\pi/\Lambda - k(\omega) - k(2\omega_0 - \omega) \ , \tag{9}$$

where $k_p = k(2\omega_0)$ and $\Lambda$ is the period of the poling in a quasi-phase-matched crystal, which compensates for the nominal mismatch $k(2\omega_0) - 2k(\omega_0)$. For collinear Type-0 or Type-1 the phase mismatch is approximated by $\Delta k(\omega) \approx -k''\cdot(\omega - \omega_0)^2$, where $k'' = \partial^2[n(\omega)\omega/c]/\partial\omega^2$ is the group-velocity dispersion. The spectral gain coefficient is denoted as

$$s(\omega) = \sqrt{\gamma^2 - \Delta k(\omega)^2/4}$$
$$\approx \sqrt{\gamma^2 - \kappa^2(\omega - \omega_0)^4} \quad , \tag{10}$$

using the abbreviation $\kappa \equiv k''/2$. The (real) gain coefficient is $\gamma = (\omega_0/c)\chi_0 E_{p0}$, where $\chi_0$ is proportional to $\chi^{(2)}$ and is assumed to be independent of frequency in the region of interest. Type-II and noncollinear Type-0 or Type-I phase matching are treated in **Appendix B**.

Note the symmetries $s(2\omega_0 - \omega) = s(\omega)$, $f(2\omega_0 - \omega) = f(\omega)$ and $g(2\omega_0 - \omega) = g(\omega)$, valid for Type-0 or Type-I phase matching. And note that unitarity of the transformation is ensured by the relation

$$|f(\omega)|^2 - |g(\omega)|^2 = 1 . \tag{11}$$

If the initial field state is the vacuum, a squeezed-vacuum state is generated; vacuum fluctuations are amplified, creating correlated pairs of photons. With a CW pump, the stationary PDC field has a spectral flux $S(\omega)$ (photons per second per frequency interval) related to the two-frequency correlation function by

$$\langle vac|\hat{b}^\dagger(\omega)\hat{b}(\omega')|vac\rangle = S(\omega)2\pi\delta(\omega - \omega') \tag{12}$$

and

$$S(\omega) = |g(\omega)|^2 = \gamma^2 \left|\frac{\sinh[s(\omega)z]}{s(\omega)}\right|^2 . \tag{13}$$

Plots of the spectrum are shown in **Fig. 3**.

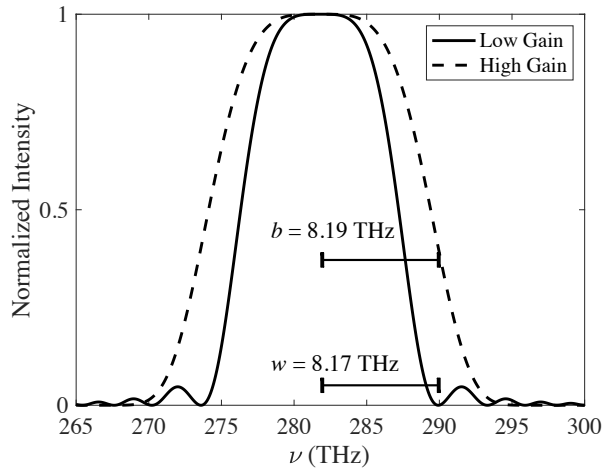

**Fig. 3**. PDC spectra in the low- and high-gain regimes, with characteristic widths $w$ and $b$, respectively. In the low-gain regime, the spectrum is well approximated by $\text{sinc}^2[(\omega-\omega_0)^2]$ and $w$ is defined by its first zero-crossing. In the high-gain regime, the spectrum is well approximated by a super-gaussian as in Eq.(19), and the $b$ is defined by the $e^{-1}$ crossing. Crystal length $z = 0.01\, m$. Low-gain: $\gamma z = 10^{-4}$, High-gain: $\gamma z = 10$.

The total photon rate [photons s$^{-1}$] is given by the integrated spectrum,

$$F = \int |g(\omega)|^2 \, d\omega \,. \tag{14}$$

In the low-gain limit ($\gamma \to 0$) the spectrum becomes [32]

$$S(\omega) \approx (\gamma z)^2 \left( \frac{\sin\left(\kappa(\omega-\omega_0)^2 z\right)}{\kappa(\omega-\omega_0)^2 z} \right)^2 \,. \tag{15}$$

The characteristic width parameter for this low-gain spectrum is [rad/s]

$$w = \sqrt{\pi/\kappa z} \,, \tag{16}$$

which narrows slowly as the medium length $z$ increases. The full width at half maximum in rad/s is $2\sqrt{1.39/\kappa z} = 2\sqrt{1.39/\pi}\, w \approx 1.34 w$. The total photon rate in this case is

$$F^{low\, gain} = (2/3\pi)(\gamma z)^2 w \,. \tag{17}$$

For later reference, the forms of $s(\omega)$, $f(\omega)$ and $g(\omega)$ in the low-gain limit are,

$$\begin{aligned} s(\omega) &\to i\kappa(\omega-\omega_0)^2 \\ f(\omega) &\to \exp[i\kappa(\omega-\omega_0)^2 z] \\ g(\omega) &\to i\gamma(\omega) z \frac{\sin[\kappa(\omega-\omega_0)^2 z]}{\kappa(\omega-\omega_0)^2 z} \end{aligned} \tag{18}$$

Referring to Eq.(7), we see that in the low-gain limit $f(\omega)$ represents linear dispersion of the input field (which in our case is vacuum) and $g(\omega)$ represents the lowest order of photon pair generation as determined by phase matching.

In the high-gain limit ($\gamma z \gg 1$) the spectrum becomes 'super-gaussian', [32]

$$S(\omega) \approx \frac{1}{4}\exp(2\gamma z)\exp\left[-(\omega-\omega_0)^4/b^4\right] , \tag{19}$$

with a width parameter

$$b = (\gamma / \kappa^2 z)^{1/4}$$
$$= (\gamma z / \pi^2)^{1/4} w \quad . \quad (20)$$

The half width at half maximum is $(\gamma \ln(2)/\kappa^2 z)^{1/4} = b(\ln(2))^{1/4} \approx 0.91 b$. The high-gain approximation holds inside the center spectral region defined as $|\omega - \omega_0| < (\gamma/\kappa)^{1/2} = b(\gamma z)^{1/4}$, which fully contains most of the energy.

The total photon rate in this case is (using the gamma function, $\Gamma(5/4) \approx 0.9064$) [32]

$$F^{high\ gain} \approx 0.91 (b/4\pi) \exp(2\gamma z) \quad . \quad (21)$$

The growth of the total intensity is nearly exponential in gain and in medium length, altered slightly by the bandwidth factor $b$.

For later reference, the forms of $s(\omega)$, $f(\omega)$ and $g(\omega)$ in the high-gain limit are

$$s(\omega) \to \gamma$$
$$f(\omega) \to \frac{1}{2} \exp[\gamma z] \exp\left[-\left(\frac{\kappa^2 z}{2\gamma}\right)(\omega - \omega_0)^4\right] \quad . \quad (22)$$
$$g(\omega) \to i f(\omega)$$

## 4. Temporally gated squeezed field

To model the interaction of the CW squeezed field with the molecule for a finite time, we impose the action of a sudden open-or-closed temporal gate, which multiplies the field by a function $\tilde{W}(t)$ that equals 1 inside the window $\{-T/2, T/2\}$ and zero otherwise. Initially we assume that higher-order linear dispersion within or subsequent to the PDC crystal is minimal and need not be compensated by use of an adjustable dispersive delay line. Compensation of dispersion is treated in **Section 11**. The time gate creates a 'rectangular' pulse of otherwise stationary squeezed light, which is simpler to handle theoretically than a squeezed field created by a pulsed pump field, as considered in [7] or [18]. The gate function in the frequency domain is

$$W(\omega) = \int_{-\infty}^{\infty} \tilde{W}(t) e^{i\omega t} dt$$
$$= T \, \text{sinc}[\omega T / 2] \quad . \quad (23)$$

The temporal gating action leads to a convolved operator

$$\hat{c}(\omega) = \int d\omega' W(\omega - \omega') \hat{b}(\omega')$$
$$\approx f(\omega) \int d\omega' W(\omega - \omega') \hat{a}(\omega') + g(\omega) \int d\omega' W(\omega - \omega') \hat{a}^\dagger (2\omega_0 - \omega') \quad , \quad (24)$$

where in the second line we assumed that $T$ is much greater than the light's coherence time (inverse of its spectral width) and thus took the gain functions outside the integrals. (To model an ultrashort 'rectangular' pulse 'chopped' from a CW source, the analysis could be carried out without this approximation, but this case is not our focus and the result would not agree with models using a short pump pulse.) This form motivates defining filtered creation and annihilation operators

$$\hat{A}(\omega) = \int d\omega'' W(\omega - \omega'') \hat{a}(\omega'')$$
$$\hat{B}^\dagger(\omega) = \int d\omega'' W(\omega - \omega'') \hat{a}^\dagger (2\omega_0 - \omega'') \quad , \quad (25)$$

so,

$$\hat{c}(\omega) = f(\omega) \hat{A}(\omega) + g(\omega) \hat{B}^\dagger(\omega) \quad . \quad (26)$$

Strictly speaking, we should include extra additive terms in Eq.(25) to account for the field operators that impinge on the temporal gate during times in which it is closed. These Langevin 'vacuum-field noise terms' would ensure unitarity such that the commutator of $\hat{A}(\omega)$ and $\hat{A}^\dagger(\omega')$ would be $2\pi\delta(\omega-\omega')$. We can omit those extra terms here because they do not contribute to detectable photons nor to excitation of the molecules. Without including these extra terms, the commutator is found to be spectrally and temporally broadened. Denoting it by $D(\omega - \tilde{\omega})$, it is given by

$$[\hat{A}(\omega), \hat{A}^\dagger(\tilde{\omega})] \equiv D(\omega - \tilde{\omega})$$
$$= \int d\omega' W(\omega - \omega') W(\tilde{\omega} - \omega') \quad (27)$$
$$= T \operatorname{sinc}[(\omega - \tilde{\omega})T/2]$$

which is normalized as $D(0) = T$ and

$$\int d\omega D(\omega - \tilde{\omega}) = 1,$$
$$\int d\omega D(\omega - \tilde{\omega})^2 = T \quad (28)$$

and acts like a (fat) delta function when multiplying broader functions such as $f(\omega)$ and $g(\omega)$.

Because $W(\omega)$ is symmetric, we have $\hat{B}(\omega) = \hat{A}(2\omega_0 - \omega)$ and thus

$$\hat{c}(\omega) = f(\omega)\hat{A}(\omega) + g(\omega)\hat{A}^\dagger(2\omega_0 - \omega) \quad , \quad (29)$$

which is a two-mode Bogoliubov (squeezing) transformation involving frequencies symmetrically displaced from the center frequency. [33] Then, using Eq.(10), one can show that

$$[\hat{c}(\omega), \hat{c}^\dagger(\omega')] = \left[|f(\omega)|^2 - |g(\omega)|^2\right] D(\omega - \omega')$$
$$= D(\omega - \omega') \tag{30}$$

consistent with $\hat{c}(\omega)$ being a filtered field operator.

Starting in the time domain, one can prove, using Parseval's theorem, that the mean number of photons in the gated squeezed-light pulse of duration $T$ equals

$$N = \int d\omega \langle vac|\hat{c}^\dagger(\omega)\hat{c}(\omega)|vac\rangle$$
$$= T \int d\omega |g(\omega)|^2 \tag{31}$$

consistent with the form of the mean photon rate (photons s$^{-1}$) given in Eq.(14).

Regarding the number of photons in a pulse, there are three regimes of interest. 1) Ultralow flux PDC wherein the whole pulse contains one or fewer photon pairs, 2) Intermediate flux wherein the whole pulse contains many photon pairs but each field mode has less than one photon as a consequence of low squeezing gain, and 3) High flux wherein the each field mode contains many photon pairs as a consequence of high squeezing gain. To quantify these regimes, we note that the effective number of temporal-spectral modes $M$ is equal to the time-bandwidth product, $M = BT$, where $B$ is the full bandwidth in Hz of the squeezed field (equal approximately to $1.34w/2\pi$ as given by Eq.(16)). Thus, the mean number of photons per temporal mode, denoted $\bar{n}_{est}$, is estimated as

$$\bar{n}_{est} \approx \frac{N}{M} = \frac{N}{BT} = \frac{F}{B}, \tag{32}$$

where $F$ is the photon rate [photons s$^{-1}$] in Eq.(14). This form can be understood as the mean number of photons per coherence time $1/B$.

**5. Four-frequency correlation function**

The four-frequency correlation function Eq.(5), needed to calculate $g^{(2)}(0)$ and the TPA rate, can be expressed, using $|\Psi\rangle = |vac\rangle$,

$$C^{(4)}(\omega_a, \omega_b, \omega_c, \omega_d) = \langle vac|\hat{c}^\dagger(\omega_a)\hat{c}^\dagger(\omega_b)\hat{c}(\omega_c)\hat{c}(\omega_d)|vac\rangle$$
$$\equiv \langle \varphi(\omega_a, \omega_b)|\varphi(\omega_c, \omega_d)\rangle \tag{33}$$

where

$$|\varphi(\omega_c,\omega_d)\rangle = \hat{c}(\omega_c)\hat{c}(\omega_d)|vac\rangle \\ = |\varphi_1(\omega_c,\omega_d)\rangle + |\varphi_2(\omega_c,\omega_d)\rangle \tag{34}$$

where

$$|\varphi_1\rangle = f(\omega_c)g(\omega_d)D(\omega_c+\omega_d-2\omega_0)|vac\rangle \\ |\varphi_2\rangle = g(\omega_c)g(\omega_d)\hat{A}^\dagger(2\omega_0-\omega_c)\hat{A}^\dagger(2\omega_0-\omega_d)|vac\rangle \tag{35}$$

To derive the $|\varphi_1\rangle$ result we used, from Eq.(27),

$$\hat{A}(\omega_c)\hat{A}^\dagger(\omega_a)|vac\rangle \equiv D(\omega_c-\omega_a)|vac\rangle \tag{36}$$

Because $|\varphi_1\rangle$ and $|\varphi_2\rangle$ are orthogonal, we have for the four-frequency correlation function $C^{(4)}(\omega_a,\omega_b,\omega_c,\omega_d) = C_{coh} + C_{incoh}$, where

$$C_{coh} = \langle \varphi_1(\omega_a,\omega_b)|\varphi_1(\omega_c,\omega_d)\rangle \\ = f^*(\omega_b)g^*(\omega_a)D(\omega_b+\omega_a-2\omega_0)f(\omega_c)g(\omega_d)D(\omega_c+\omega_d-2\omega_0) \\ \approx f^*(2\omega_0-\omega_a)g^*(\omega_a)f(\omega_c)g(2\omega_0-\omega_c) \times \\ D(\omega_b+\omega_a-2\omega_0)D(\omega_c+\omega_d-2\omega_0) \tag{37}$$

and

$$C_{incoh} = \langle \varphi_2(\omega_a,\omega_b)|\varphi_2(\omega_c,\omega_d)\rangle \\ = g^*(\omega_b)g^*(\omega_a)g(\omega_c)g(\omega_d) \times \\ \langle vac|\hat{A}(2\omega_0-\omega_b)\hat{A}(2\omega_0-\omega_a)\hat{A}^\dagger(2\omega_0-\omega_c)\hat{A}^\dagger(2\omega_0-\omega_d)|vac\rangle \\ = g^*(\omega_b)g^*(\omega_a)g(\omega_c)g(\omega_d) \times \\ D(\omega_c-\omega_a)D(\omega_b-\omega_d) + D(\omega_c-\omega_b)D(\omega_a-\omega_d) \\ \approx |g(\omega_c)|^2|g(\omega_d)|^2\left(D(\omega_c-\omega_a)D(\omega_b-\omega_d)+\xi D(\omega_c-\omega_b)D(\omega_a-\omega_d)\right) \tag{38}$$

In deriving these results we swapped $\omega_a,\omega_b$ and used $\omega_b \to 2\omega_0-\omega_a$, $\omega_d \to 2\omega_0-\omega_c$, along with the fact that $f(\omega)$ and $g(\omega)$ are assumed to be broad compared to the $D$ functions (for large $T$).

The sum of the products of $D$ functions in $C_{incoh}$ arises from the nonzero commutator of $\hat{A}(\omega)$ and $\hat{A}^\dagger(\tilde{\omega})$ in Eq.(27). Here we note (see **Appendix B** for proof) that for Type-II or off-axis Type-0 or Type-I phase matching, the result for $C_{incoh}$ lacks the second product of $D$ functions, because in those cases $\hat{A}(\omega)$ and $\hat{A}^\dagger(\omega)$ are labeled by distinguishing indexes, so the relevant commutator equals zero. In the above we therefore inserted a 'flag' $\xi$, which equals 1 for the indistinguishable cases (colinear, co-polarized Type-0 or Type-I) and equals zero for the distinguishable cases. We show in **Section 6** that when the squeezed-light bandwidth is large compared to the TPA linewidth the term $C_{incoh}$ contributes negligibly to the TPA rate. In the opposite case it does make a significant contribution, as we show in **Section 9**.

## 6. TPA by weak or bright squeezed vacuum

The probability for two-photon excitation of the final molecular state following the time-gated squeezed-state pulse is evaluated, using Eq.(3), as $P^{TPA} = P_{coh} + P_{incoh}$, where

$$
\begin{aligned}
P_{coh} &= \sigma^{(2)} \frac{\gamma_{fg}}{A_0^2} \text{Re} \int d\omega' \int d\omega \int d\tilde{\omega} \frac{C_{coh}(\omega',\tilde{\omega}',\omega,\tilde{\omega})}{\gamma_{fg} - i\omega_{fg} + i\omega + i\tilde{\omega}} \\
P_{incoh} &= \sigma^{(2)} \frac{\gamma_{fg}}{A_0^2} \text{Re} \int d\omega' \int d\omega \int d\tilde{\omega} \frac{C_{incoh}(\omega',\tilde{\omega}',\omega,\tilde{\omega})}{\gamma_{fg} - i\omega_{fg} + i\omega + i\tilde{\omega}}
\end{aligned}
\quad (39)
$$

Using the forms of the correlation functions, noting again that $f(\omega)$ and $g(\omega)$ are broad compared to the $D$ functions for sufficiently large $T$, using the symmetry $g(2\omega_0 - \omega) = g(\omega)$, and inserting $\tilde{\omega}' = \omega + \tilde{\omega} - \omega'$, we find

$$
\begin{aligned}
P_{coh} &= \sigma^{(2)} \frac{\gamma_{fg}}{A_0^2} \text{Re} \int d\omega' \int d\omega \int d\tilde{\omega} \\
&\quad \times \frac{f^*(2\omega_0 - \omega')g^*(\omega')f(\omega)g(2\omega_0 - \omega)D(\omega + \tilde{\omega} - 2\omega_0)D(\omega + \tilde{\omega} - 2\omega_0)}{\gamma_{fg} - i\omega_{fg} + i\omega + i\tilde{\omega}} \\
&\approx \frac{\sigma^{(2)} T}{A_0^2} \frac{\gamma_{fg}^2}{\gamma_{fg}^2 + (\omega_{fg} - 2\omega_0)^2} \left| \int d\omega f(\omega) g(\omega) \right|^2
\end{aligned}
\quad (40)
$$

and

$$
\begin{aligned}
P_{incoh} &= \sigma^{(2)} \frac{\gamma_{fg}}{A_0^2} \text{Re} \int d\omega' \int d\omega \int d\tilde{\omega} \frac{|g(\omega)|^2 |g(\tilde{\omega})|^2 \left( D^2(\omega' - \omega) + \xi D^2(\omega' - \tilde{\omega}) \right)}{\gamma_{fg} - i\omega_{fg} + i\omega + i\tilde{\omega}} \\
&\approx (1+\xi) \frac{\sigma^{(2)} T}{A_0^2} \int d\omega \int d\tilde{\omega} \frac{\gamma_{fg}^2}{\gamma_{fg}^2 + (\omega_{fg} - \omega - \tilde{\omega})^2} |g(\omega)|^2 |g(\tilde{\omega})|^2
\end{aligned}
\quad (41)
$$

We can see that in the low-gain regime, where $f(\omega) \approx 1$, the coherent contribution $P_{coh}$ scales linearly in the photon flux $|g(\omega)|^2$, whereas $P_{incoh}$ scales quadratically in the photon flux $|g(\omega)|^4$. And in the high-gain regime, where $f(\omega) \approx g(\omega)$, both contributions scale quadratically. It is worth noting that the coherent and incoherent contributions play similar roles in sum-frequency generation, as studied theoretically and experimentally. [18, [34], [35]]

While the foregoing expressions can be evaluated generally to account for the spectral overlap of the squeezed light and the molecular absorption profile, we first focus on the case where the exciting squeezed field has a much broader bandwidth than the two-photon transition linewidth and twice its center frequency is resonant with the two-photon transition. In this case, the frequency anticorrelation between photons within a pair makes their combined action act as if monochromatic light is driving the TPA. We call this effect *spectral compression* and discussed it in detail in [1]. In this case we have, because the Lorentzian acts like a delta function,

$$P_{coh} \approx \frac{\sigma^{(2)}T}{A_0^2}\left|\int \dbar\omega f(\omega)g(\omega)\right|^2$$

$$P_{incoh} \approx \frac{1+\xi}{2}\frac{\sigma^{(2)}T}{A_0^2}\gamma_{fg}\int \dbar\omega |g(\omega)|^4$$

(42)

In **Fig. 4** we show plots of the coherent and incoherent contributions to the TPA probability using Eq.(42) (within its regimes of validity) and realistic parameters for a typical dye molecule for large squeezed-light bandwidth large compared to the linewidth of the molecular transition.

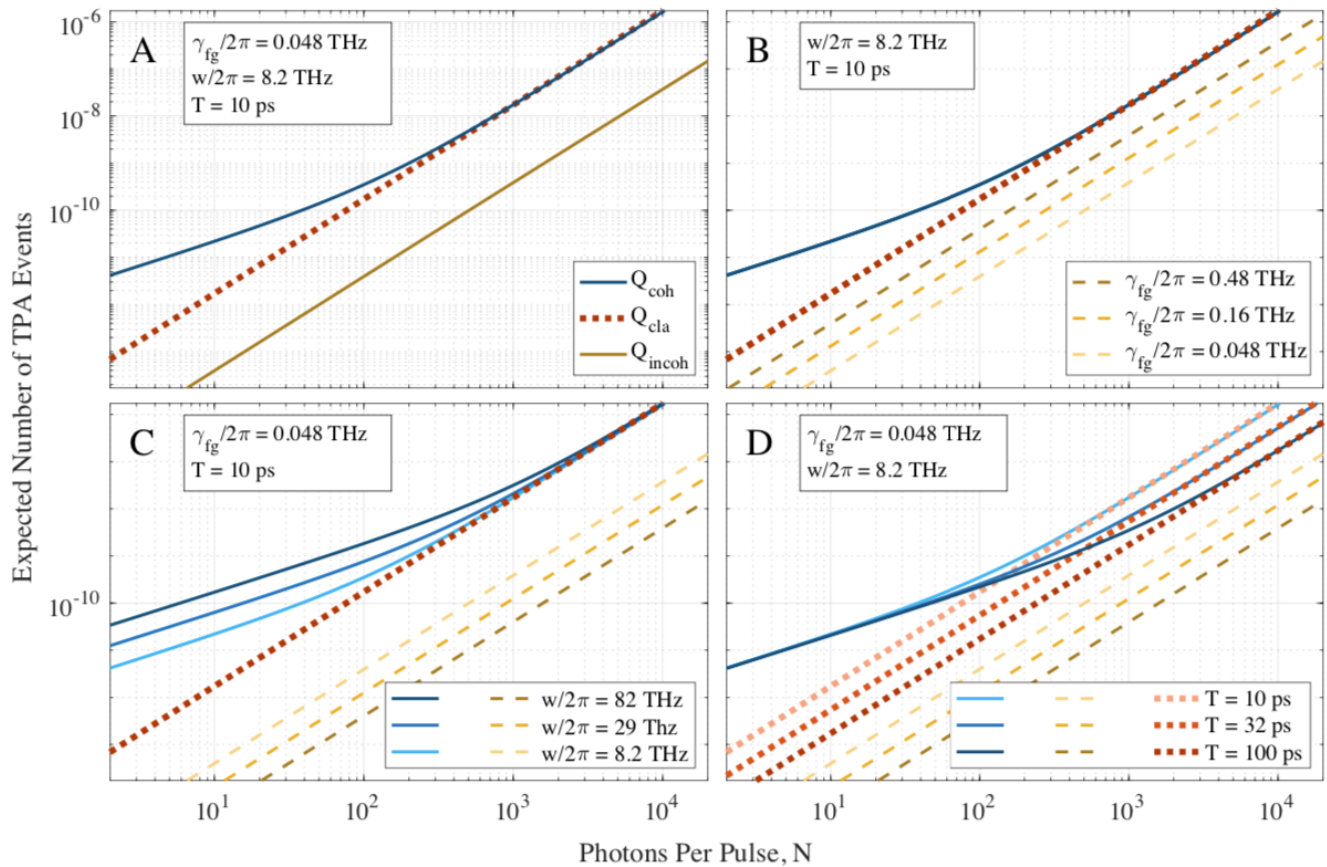

Fig. 4 Predicted mean number of molecules excited by TPA per pulse for a final-state TPA linewidth that is much narrower than the BSV bandwidth but broader than the effective PDC pump bandwidth, from Eq. (42). Realistic experimental parameters are: 10 μm effective beam radius (assumed collimated), 1 cm cuvette, and 10 mmol concentration of molecules assumed to have 9 GM TPA cross section. TPA probability per molecule is evaluated from Eq.(42), assuming that twice the squeezed-light center frequency is resonant with the two-photon transition. In each case the solid blue curve is the coherent ETPA contribution, the dashed yellow curve is the incoherent ETPA contribution, and the dotted red line is 'classical' coherent-state TPA with a quasi-monochromatic pulse duration $T$ that matches that of the laser pumping the PDC process. Part A) shows a representative case. A transition in scaling from linear in photon number to quadratic is apparent at $N \sim 125$, which corresponds to a mean occupation of 1 photon per temporal mode. B) shows the effect of changes in the absorber's linewidth, which affects only the incoherent contribution. For this plot we held $\sigma^{(2)}$ constant, which implies from Eq.(4) that the dipole strengths are varied to compensate for varying $\gamma_{fg}$. C) shows the effect of increasing the low-gain bandwidth parameter $w$ of the squeezed light, varied by varying the crystal length $z$. The TPA efficiency of the coherent contribution is increased in the low-gain regime and remains the same in the high-gain regime, while the incoherent contribution decreases in efficiency. D) shows the effect of increasing the time window $T$. In the low-gain regime the coherent contribution remains unchanged; however, the high-gain efficiency is reduced by increasing $T$, and the crossover to quadratic scaling occurs at a higher relative photon number. Both incoherent and classical efficiency are reduced.

Note that in the low-gain regime the excitation probability is independent of pulse duration $T$ for fixed $N$. That is because the entangled photons arrive in tight pairs regardless of the arrival times of each pair. The crossover from linear to quadratic scaling is evident, consistent with Eq.(1) and known from prior studies.

Also plotted in **Fig. 4** (as the red dashed line) is the prediction for excitation by a quasi-monochromatic coherent state, using Eq.(112) from [15], valid for a 'rectangular' classical-light pulse with duration $T$ much longer than the inverse linewidth of the absorber. In this case the probability is

$$P_{coherent\ state} = \frac{\sigma^{(2)}}{A_0^2}\left(\frac{N}{T}\right)^2 T . \tag{43}$$

Note that Eq.(40) predicts a TPA probability versus pump frequency that is Lorentzian and has the same linewidth as the molecular transition, which can be much narrower than the bandwidth of the BSV. Such a narrow TPA spectrum was observed in the experiment by Dayan, et. al. [36]

Crucially, in the high-gain regime the quasi-monochromatic coherent-state result is the same as the squeezed-state result. Thus, a major conclusion of the present study is that broad-band squeezing gives no advantage in TPA rate compared to a quasi-monochromatic coherent state pulse of the same duration and energy. The only difference is that in broad-band squeezing the spectral density of light is spread over a wide bandwidth and contains frequency correlations, while in the coherent state the light is concentrated in a near-monochromatic spectrum.

## 7. Analytical rate expression in broad-band limit

A central result of the present study is a new analytical formula for the rate of TPA in the case that the TPA linewidth is much narrower than the squeezed-light bandwidth. We find, as derived later in this section, for the ETPA probability per molecule,

$$P_{coh} \approx \left(\frac{N}{A_0 T}\right)\frac{\sigma^{(2)}}{A_0}\frac{3}{4}\frac{w}{\pi}T + \sigma^{(2)}\left(\frac{N}{A_0 T}\right)^2 T, \tag{44}$$

which is consistent with Eq.(1) and Eq.(2) upon identifying the flux of squeezed light as $F = N/T$ and the effective bandwidth in Hz as $B/2\pi \approx 3w/4\pi$ (depending on the convention used to define bandwidth), where $w = \sqrt{\pi/\kappa z}$ is the squeezed-light bandwidth in the low-gain regime, from Eq.(16).

We plot a quantity proportional to that in Eq.(44) in **Fig. 5**, showing good agreement with the numerical evaluation.

The crossover between linear and quadratic scaling of the TPA rate with flux is found by equating the two terms on the right hand side of Eq. (44). This gives

$$N_{cross\ over} = \frac{3}{4\pi} wT \ . \qquad (45)$$

The crossover occurs when the number of photons *per mode* begins to exceed roughly one, in agreement with previous studies. [4, 5, 7,18]

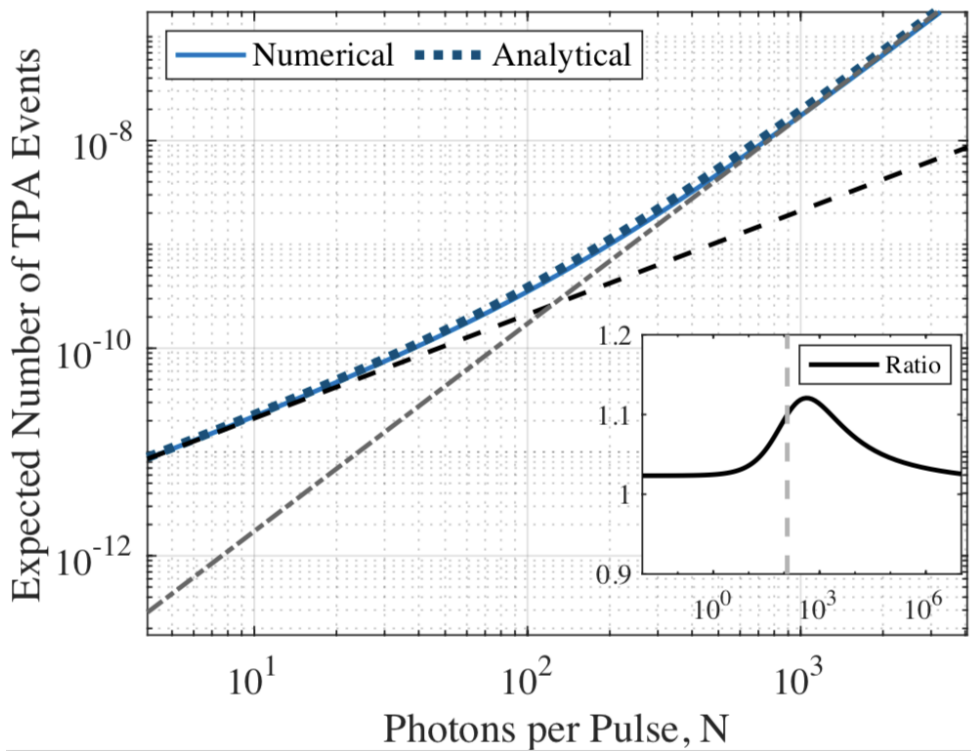

Fig. 5. Predicted mean number of molecules excited by TPA per pulse given realistic experimental parameters: 10 µm beam radius, 1 cm cuvette length, 9G M TPA cross section, and 10 mmol concentration, for numerical evaluation of Eq.(42) (solid blue curve), as well as the analytical expression given in Eq.(44) (dotted dark blue curve). The analytical expression for TPA probability is in good agreement with the numerical results, limited primary by the accuracy of the approximation of the width function. The black dashed line shows the low-gain limiting behavior and the dash-dotted dark-grey line shows the high-gain limiting behavior. The inset shows the ratio of the analytical function and the numerical results over a wide range of values. The maximum deviation is near the crossover point (In the inset the light-grey dashed line) and is within 15% of the numerical value. No dispersion compensation has been applied.

In **Appendix D** we show that the discrepancy between numerical and our approximate form can be reduced from about 15% as seen in **Fig. 5** to less than 3% by including an optimum amount of dispersion compensation subsequent to the PDC crystal.

We derived Eq.(44) as follows. First consider the low-gain limit of the coherent term

$$P_{coh}^{low\,gain} = \frac{\sigma^{(2)}N}{A_0^2} \frac{\left|\int d\omega f(\omega)g(\omega)\right|^2}{\int d\omega |g(\omega)|^2}$$
$$\approx \frac{\sigma^{(2)}N}{A_0^2} \frac{\left|\int d\omega g(\omega)\right|^2}{\int d\omega |g(\omega)|^2} = \frac{\sigma^{(2)}N}{A_0^2} \frac{3}{4\pi} w \quad . \tag{46}$$

To arrive at this result we used the low-gain expressions for $f(\omega)$ and $g(\omega)$ in Eq.(18) and replaced $f(\omega)$ by 1 under the assumption that the linear (second-order group delay) dispersion of the squeezed field has been removed by use of a pulse compressor-like dispersive delay line. [37]

Next consider the high-gain limit of the coherent term

$$P_{coh}^{high\,gain} = \sigma^{(2)} \frac{N^2}{A_0^2 T} \mu^2 \quad , \tag{47}$$

where

$$\mu = \frac{\left|\int d\omega f(\omega)g(\omega)\right|}{\left(\int d\omega |g(\omega)|^2\right)} \approx 1 \quad , \tag{48}$$

and we used the high-gain expressions for $f(\omega)$ and $g(\omega)$ in Eq.(22) and carried out the integrals. To obtain the main result Eq.(44) we simply sum the low- and high-gain expressions for $P_{coh}$, since one or the other dominates in the two regimes of interest.

It remains to show that the incoherent terms, which scale as $N^2$, are negligible compared to the coherent terms. This term gives a rate per molecule

$$R_{incoh} = \frac{P_{incoh}}{T} = \frac{1+\xi}{2} \sigma^{(2)} \left(\frac{N}{A_0 T}\right)^2 \beta \quad , \tag{49}$$

where

$$\beta = \gamma_{fg} \frac{\int d\omega |g(\omega)|^4}{\left(\int d\omega |g(\omega)|^2\right)^2} \approx \begin{cases} \gamma_{fg} 1.26\sqrt{\pi\kappa z} = \gamma_{fg} 1.26\frac{\pi}{w}, & \text{low gain} \\ \gamma_{fg} 1.10\pi(\kappa^2 z/\gamma)^{1/4} = \gamma_{fg} 1.10\frac{\pi}{b}, & \text{high gain} \end{cases}, \quad (50)$$

where we again used Eq.(18) for the low-gain expressions assuming dispersion is compensated (See **Section 10**), and Eq.(22) for the high-gain expressions, carried out the integrals, and identified the bandwidth $b = (\gamma/\kappa^2 z)^{1/4}$ in the high-gain limit (that is $\gamma z \gg 1$) from Eq.(20). We need only compare this expression to the $N^2$ scaling coherent term in Eq.(44), that is $\sigma^{(2)}(N/A_0 T)^2$. Indeed, the factor $\beta$ is much less than 1 in both regimes, because the bandwidth of the squeezed field is assumed here to be much greater than the molecular linewidth, $w, b \gg \gamma_{fg}$.

## 8. Second-order intensity autocorrelation function $g^{(2)}(0)$

As a precursor to deriving the TPA rate in the case that the squeezed-light bandwidth is small compared to the TPA linewidth, and thus the molecular response to the intensity is near-instantaneous, we calculate the second-order intensity autocorrelation function at zero time delay. It is found, using the four-frequency correlation function, to be (see **Appendix D**)

$$g^{(2)}(0) = \frac{\langle \hat{E}^{(-)}(0)\hat{E}^{(-)}(0)\hat{E}^{(+)}(0)\hat{E}^{(+)}(0) \rangle}{\langle \hat{E}^{(-)}(0)\hat{E}^{(+)}(0) \rangle^2}$$

$$= (1+\xi) + \frac{\left|\int d\omega f(\omega) g(\omega) \exp[i(\omega-\omega_0)^2 D_2]\right|^2}{\left(\int d\omega |g(\omega)|^2\right)^2}, \quad (51)$$

where we inserted a factor to represent a dispersion-compensating device inserted as in **Fig. 2**, with $D_2$ being its second-order (group delay) dispersion. (See **Section 10**.) A similar result is found in [31] but without detailed evaluation or consideration of dispersion compensation. In **Appendix D** we show that to good approximation $g^{(2)}(0)$ can be written, by using the $f(\omega)$ and $g(\omega)$ solutions given above as

$$g^{(2)}(0) = (2+\xi) + \frac{1}{\bar{n}}, \quad (52)$$

where $\bar{n}$ is the mean number of photons per mode,

$$\bar{n} = \frac{N}{(3w/4\pi)T} = \frac{F}{(3w/4\pi)}, \tag{53}$$

which is independent of the time-gate duration $T$, as expected. This result can be seen to be consistent with the mean number of photons per temporal mode in Eq.**(32)**. Recalling the full bandwidth at half maximum in terms of $w$ following Eq.**(16)**, we see the two forms are in good agreement because $3w/4\pi \approx 1.34 w/2\pi$, the same within 12%. (The particular value depends on the functional form of the squeezed-light spectrum.)

For Type-2 PDC or non-collinear Type-0 or 1 PDC, where photons are distinguishable ($\xi = 0$), we have $g^{(2)}(0) = 2 + 1/\bar{n}$. For collinear Type-0 or 1 PDC, where photons are indistinguishable ($\xi = 1$), this result reproduces the result known in idealized single-mode squeezing theory, $g^{(2)}(0) = 3 + 1/\bar{n}$. [22, 23] It is worth noting that the form Eq.(52) can be obtained only when compensating the dispersion optimally at each value of parametric gain, to maximize the magnitude of the instantaneous intensity fluctuations. Without such compensation it is found that $g^{(2)}(0)$ drops below the value $(2 + \xi)$ for intermediate values of $\bar{n}$.

These results are illustrated in **Fig.6**, where we plot the compensated and uncompensated forms of $g^{(2)}(0)$ versus the mean total photon number per pulse $N = T \int |g(\omega)|^2 d\omega$. The uncompensated $g^{(2)}(0)$ dips significantly below 3 because dispersion in the PDC crystal stretches the photon wave packets reducing their peak fluctuation intensities. For the compensated form, we optimize numerically the value of $D_2$ at each value of photon number to maximize the magnitude of the instantaneous intensity fluctuations. These results will be important when we consider its relation to two-photon absorption.

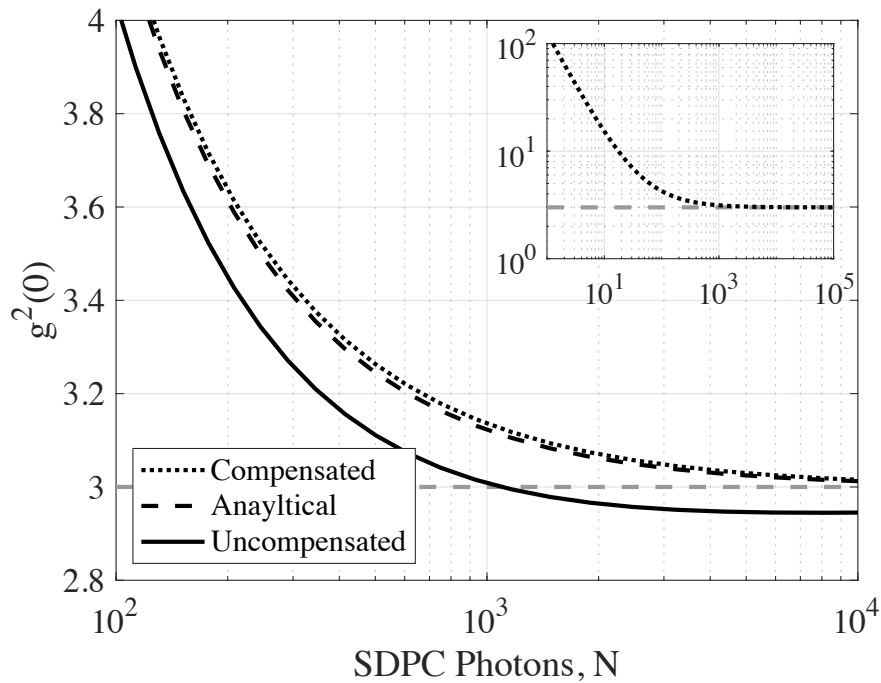

**Fig. 6** Log-Linear plot of $g^{(2)}(0)$ as a function of the mean number of photons per pulse of squeezed light, plotted for differing assumptions regarding the value of dispersion $D_2$, with remaining parameters held constant. 'Uncompensated' is calculated numerically and assumes no dispersion compensation. 'Compensated' is calculated numerically, and uses optimal second-order dispersion compensation, also found numerically. 'Analytical' assumes the analytical model, $g^{(2)}(0) = 3 + 1/\bar{n}$. Inset: Log-Log plot of optimally compensated curve with the same axes as the main figure. 1/n scaling can be seen in the low-gain regime in good agreement with theoretical predictions. For these plots we assumed $T = 10$ ps, although both axes simply scale linearly with value of $T$.

We can understand why the result found here from explicit calculation of our multi-spectral-mode BSV model agrees with the idealized single-mode result $g^{(2)}(0) = 3 + 1/\bar{n}$. While the single-mode result is often understood as single-spectral mode, the same holds for single-temporal mode. For instantaneous detection, as modeled by g⁽²⁾(0) or ultrafast detection in a window shorter than a coherence time 1/*B*, the detected field is effectively filtered to a single temporal mode.

## 9. Transition from broad-band to narrow-band excitation of TPA MOVED HERE

Here we treat the transition to the case that the squeezed-light bandwidth is much narrower than the TPA molecular linewidth, wherein the molecules respond instantaneously to fluctuations of the light. We confirm and generalize the known result that in this limit, the TPA rate is proportional to the second-order (intensity) correlation function g$^{(2)}$ (0) (when dispersion compensation is invoked), leading in the case of BSV to a ×3 enhancement of TPA relative to quasi-monochromatic coherent-state light. [17, 20, 21, 31]

From Eqs.**(40)** and **(41)** we derive in this limit,

$$P_{coh} + P_{incoh} = \frac{\sigma^{(2)}T}{A_0^2}\left(\frac{N}{T}\right)^2 \frac{\gamma_{fg}^2}{\gamma_{fg}^2 + (\omega_{fg} - 2\omega_0)^2}\left(\left(\frac{T}{N}\right)^2 \left|\int d\omega f(\omega)g(\omega)\right|^2 + (1+\xi)\right)$$

$$= \frac{\sigma^{(2)}T}{A_0^2}\left(\frac{N}{T}\right)^2 \frac{\gamma_{fg}^2}{\gamma_{fg}^2 + (\omega_{fg} - 2\omega_0)^2} \times g^{(2)}(0)$$

(54)

where we used the form of $g^{(2)}(0)$ in Eq.**(73)**. We pointed out in **Section 8** and verified in **Appendix D** that $g^{(2)}(0)$ depends on the extent of dispersion compensation applied to the squeezed light prior to interacting with the TPA sample.

Given that the proportionality of the TPA rate to $g^{(2)}(0)$ applies only for ultrabroad TPA linewidths and narrow squeezed-light bandwidths, we next study the transition to this limit, including intermediate cases. In **Fig. 7**, we plot the number of TPA events, by evaluating Eqs.(40) and (41) numerically, using the same parameters as in **Fig.4**, for varying values of the molecular final-state linewidth, holding the classical cross section $\sigma^{(2)}$ constant, as before. It is observed that in the limit of large final-state linewidth the incoherent contribution equals twice the coherent contribution, leading to a net enhancement of a factor 3, as expected.

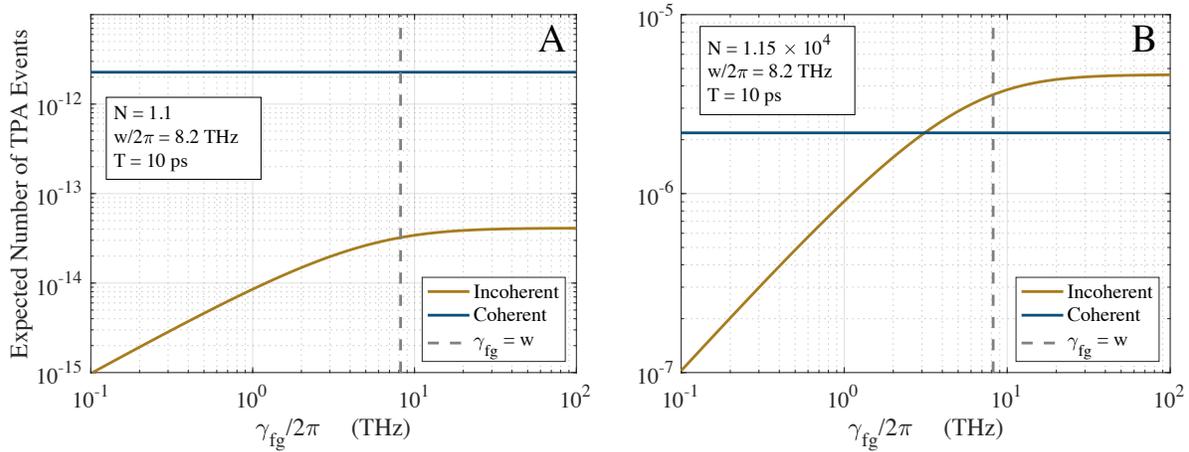

**Fig. 7** The expected number of TPA events, using the same parameters as in **Fig.4**, varying values of the molecular final-state linewidth $\gamma_{fg}$. Part (A) is a slice through Fig.4(B) at the lower end of the range of photons per pulse in that figure, showing again that for low photon flux the coherent contribution always dominates, even for larger final-state linewidths. Part (B) is a slice through Fig.4(B) at the upper end of the range of photons per pulse in that figure,

allowing larger final-state linewidths, leading to the incoherent contribution becoming comparable to or exceeding the coherent contribution. In the limit of large final-state linewidth the incoherent contribution equals twice the coherent contribution, leading to a net enhancement of a factor 3, as expected. comment on dashed line

If one wants to scale the results plotted in Figures 4 and 7 to account for different values of classical cross section $\sigma^{(2)}$, pulse duration T, and effective beam area $A_0$, simply note that all probabilities are proportional to $\sigma^{(2)}/A_0^2$. In addition, if one wants to account for inhomogeneous broadening of the molecular transition, one should integrate Eqs.(40), (41) over $\omega_{fg}$ weighted by the inhomogeneous distribution of $\omega_{fg}$ values.

## 10. Effects of Dispersion and its Compensation

It is expected that linear dispersion will decrease the TPA efficiency by spreading photon pairs in time. A question remains concerning the relative effects of dispersion in the low- and high-gain regimes and its separate effects on the coherent and incoherent contributions. To account for such effects in squeezed-light-driven TPA, we incorporate dispersive propagation into the two-photon JSA by replacing $\hat{c}(\omega)$, $f(\omega)$, and $g(\omega)$ in Eq.(26) by [15]

$$\begin{aligned}\hat{c}(\omega) &\to \hat{c}(\omega)\exp[i(D_2/2)(\omega-\omega_0)^2] \\ f(\omega) &\to f(\omega)\exp[i(D_2/2)(\omega-\omega_0)^2] \\ g(\omega) &\to g(\omega)\exp[i(D_2/2)(\omega-\omega_0)^2]\end{aligned} \quad (55)$$

in all the earlier results, where $D_2$ is the second-order (group delay) dispersion of the transmitting optical system. $D_2 > 0$ corresponds to positive chirp as caused by propagating through a typical piece of glass.

Then we see from Eq.(41) that dispersion does not affect the incoherent contribution $P_{incoh}$ for long pulses and limited dispersion. A similar result was found in [35]. From Eq.(40) we see that the coherent contribution $P_{coh}$ is reduced by a factor

$$r = \frac{\left|\int d\omega\, f(\omega)g(\omega)\exp[iD_2(\omega-\omega_0)^2]\right|^2}{\left|\int d\omega\, f(\omega)g(\omega)\right|^2} \quad (56)$$

relative to the case of no dispersion. In the low-gain case this result has been evaluated analytically using a gaussian approximation for the phase-matching function and the spectrum of the PDC pump field in [15].

We find that the amount of dispersion compensation required to optimize the TPA probability varies with pump intensity. In general, this factor must be evaluated numerically. However, the low- and high-gain approximations can be used to estimate the needed dispersion compensation for those cases. In the low-gain approximation $f(\omega) \approx \exp[i\kappa(\omega-\omega_0)^2]$, which serves as a purely dispersive factor, and can be offset with dispersion of equal magnitude and opposite sign. On the other hand, in the high-gain limit, $f(\omega) \approx g(\omega)$ with both quantities being purely imaginary, resulting asymptotically in no effective dispersion requiring compensation.

In the low-gain limit the sinc function in $g(\omega)$ of Eq.(18) has slowly decaying, oscillating tails that are unphysical far from the spectral center. Therefore we restrict the integral in Eq.(56) to the still-large frequency range $[\omega_0/2, 3\omega_0/2]$ to avoid numerical problems.

The dependence of TPA probability (from Eq.(56)) on added dispersion for compensation is plotted **Fig. 8**. For the low-gain limit we find the result that adding a small amount of negative dispersion increases the probability by about 2% for the example considered, and then, perhaps surprisingly, remains independent of further added negative dispersion until a certain threshold is reached at which point the probability begins decreasing. The boundaries of the flat-top region correspond to photon pairs generated at the entrance or exit of the PDC crystal and reflects the quantum indistinguishability of these possibilities. This behavior is not seen for the high-gain case, where most pairs are produced near the crystal exit.

We consider the effects of dispersion compensation further in **Appendix D**.

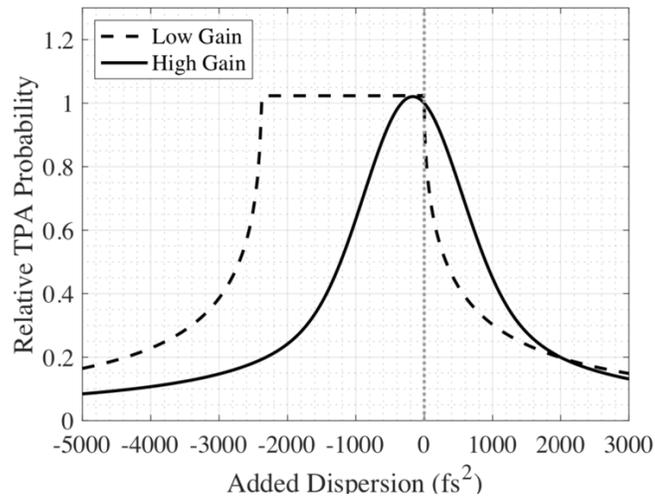

Fig. 8 Probability reduction factor $r$ versus added dispersion $D_2$, from Eq.(56). Crystal length $z = 0.01\,m$, Low-gain: $\gamma z = 10^{-4}$, High-gain: $\gamma z = 10$.

## 11. Conclusions

We presented a model for two-photon absorption of quasi-steady-state squeezed vacuum that is valid in both high- and low-gain regimes. The results, given as closed-form expressions, are evaluated easily numerically. A main finding is that if the squeezed vacuum is much broader spectrally than the molecular final-state linewidth, then bright (high-gain) squeezed vacuum is no more effective in driving TPA than is a quasi-monochromatic coherent-state (classical) pulse of the same temporal shape and mean photon number. In this case we derived analytical expressions for the TPA rate that match the numerical results well and is sufficient for explaining the transition from linear to quadratic scaling of TPA rate with PDC photon flux. Our model agrees with and generalizes previous work, is simpler to implement than numerical treatments requiring a singular-value decomposition of the PDC temporal modes, and captures the relevant dynamics for TPA driven by 'long' quasi-monochromatic pulses of squeezed light. A significant conclusion is that the sought-for advantage of observing TPA at extremely low optical flux is not provided by broadband bright squeezed vacuum.

We also considered the case of a narrow squeezed-state spectrum and an ultrabroad molecular TPA linewidth. In this case we confirmed the known result that the TPA rate is proportional to the second-order intensity autocorrelation function at zero time delay $g^{(2)}(0)$. We find that for $g^{(2)}(0)$ to reach the idealized form $g^{(2)}(0) = 3 + 1/\bar{n}$, with $\bar{n}$ being the mean number of photons per mode, dispersion compensation is required to ensure the intensity fluctuations attain their maximum values. Importantly, we also presented formulas and plots of TPA probabilities in the intermediate regime where the exciting light and the molecular TPA linewidth are comparable, showing in **Fig.7** the transition between the two limiting cases.

We can draw several conclusions about which cases lead to significant enhancement by time-frequency entanglement: A narrow two-photon absorption linewidth, coupled with a broad down-conversion bandwidth, provides maximum potential for advantage relative to a coherent laser pulse having the same bandwidth. Since the maximum classical TPA efficiency is limited by the TPA linewidth, a narrow linewidth limits the maximum classical efficiency. With a PDC pump pulse matched to this linewidth, and a broad phase-matching bandwidth, the number of photons per pulse can be large before crossing over to the high-gain regime, after which the efficiency approaches the efficiency of a classical pulse of the same temporal duration (with narrow bandwidth), and sees no large enhancement over classical light, as seen in **Fig. 4**.

Careful engineering of the PDC parameters can tune the maximum flux achievable in the low-gain regime. However, the flux at which low-gain ETPA no longer outperforms TPA from an optimal classical pulse is highly dependent on the linewidth of the two-photon transition. And notably, for typical magnitudes of the two-photon cross-section, a measurable ETPA signal in the low-gain regime remains difficult to achieve, as explained in [1, 2, 3].

Perhaps the most interesting aspect of bright-squeezed-vacuum ETPA is the ability to drive efficiently a TPA process using a field that has low spectral density at all frequencies. In contrast to the equivalent classical field, which must be narrower than the TPA linewidth to achieve optimal efficiency, the squeezed-light field can be broadband. Nevertheless, in typical cases this aspect will not serve to eliminate the optical damage that a flux high enough to create observable TPA can cause.

As argued in [17], parametric down conversion and amplification does provide a convenient method for creating light with the ability to achieve simultaneously high temporal and spectral resolution for applications in spectroscopy, which may prove to have useful benefits in spectroscopy and microscopy.

Finally, we comment that analogous effects of classical spectral correlation or quantum entanglement of driving fields can play significant roles in stimulated Raman scattering, wherein the difference of optical frequencies (rather than the sum) should be sharply defined, opposite to the case of TPA. [38, 39, 40]

**Acknowledgements**

We especially thank Maria Chekhova and Gerd Leuchs for illuminating discussions of $g^2(0)$ and John Sipe, Andy Marcus, Sofiane Merkouche, Markus Allgaier, and Brian Smith for additional helpful discussions. This work was supported by the National Science Foundation RAISE-TAQS Program (PHY-1839216).

**Data Availability**
The data supporting this study are contained within the article.

**Disclosures**
The authors declare no conflicts of interest.

**Appendix A: Theory of squeezing with indistinguishable photons**

The propagation theory for broadband squeezing has been treated many times. [30] We follow those given in [41, 32, 18, 31]. For collimated or waveguided beams in the absence of nonlinear interactions, the (vector) electric field operator within a given frequency band (spectral region with range around 5 or 10% of the carrier frequency) with center frequency $\omega_J$ is well approximated as [42, 43, 44]

$$\hat{\mathbf{E}}^{(+)}(\mathbf{r},t) = i \sum_m \int_{B_J} \frac{d\omega}{2\pi} \sqrt{\frac{\hbar \omega_J}{2 c \varepsilon_0 n_J}} \, \hat{a}_m(\omega) e^{-i\omega t} \mathbf{w}_m(\mathbf{x}) \exp(i k_m(\omega) z) \;, \tag{57}$$

where $n_J$ is the refractive index at the center frequency, and the propagation constant includes dispersion, $k_m(\omega) = \omega n_m(\omega)/c$, where $n_m(\omega)$ may be considered an effective refractive index to account for modal dispersion (in a wave guide) as well as material dispersion. The spatial mode functions are orthogonal in the transverse plane, $\int d^2x \, \mathbf{w}_n^*(\mathbf{x}) \cdot \mathbf{w}_m(\mathbf{x}) = \delta_{nm}$.

For collinear Type-0 or Type-1 PDC, the photons are indistinguishable except for their frequencies, and the distinguishing $m$ subscripts should be dropped. We write in a one-dimensional approximation

$$\hat{E}^{(+)}(z,t) \approx L_0 \int d\omega\, \hat{a}(z,\omega) e^{-i\omega t}, \tag{58}$$

where $L_0 = \hbar\omega_0 / 2c\varepsilon_0 n A_0$, $\omega_0$ is the center frequency of the down converted light, and $A_0$ is the effective beam area. To evaluate the field amplitude at an off-axis point $\mathbf{r}$, the effective beam area is replaced by $(1/A_0)^{1/2} \equiv u(\mathbf{r})$, where the mode amplitude $u(\mathbf{r})$ (which $\mathbf{w}_m(\mathbf{x})$ is proportional to) is normalized in the transverse spatial coordinates, $\int |u(\mathbf{r})|^2 d^2x = 1$. [15] We have absorbed the spatial propagation into the definition of $\hat{a}(z,\omega)$. The operator evolution in this case was formulated in the 1990s [32], which we summarize here, with a few updates, including quasi-phase matching.

The Maxwell-Heisenberg equation of motion, expressed in the frequency domain, is

$$\left[\frac{\partial}{\partial z} - ik(\omega)\right]\hat{a}(z,\omega) = i\frac{\omega_0}{c}\int d\omega'\, \chi(z)\tilde{E}_p(z,2\omega')\hat{a}^\dagger(z,2\omega'-\omega), \tag{59}$$

where $k(\omega) = k_0 + (\omega-\omega_0)k' + (\omega-\omega_0)^2 k''/2$ with $k'$ and $k''$ being first and second derivatives of $k(\omega)$ and $\tilde{E}_p(z,2\omega')$ is the Fourier transform of the pump pulse. For a monochromatic pump, define $\tilde{E}_p(z,2\omega') = E_{p0} e^{ik_p z} 2\pi\delta(\omega'-\omega_0)$ and the equation becomes:

$$\left[\frac{\partial}{\partial z} - ik(\omega)\right]\hat{a}(z,\omega) = i\frac{\omega_0}{c}\chi(z)E_{p0}e^{ik_p z}\hat{a}^\dagger(z,2\omega_0-\omega). \tag{60}$$

Quasi-phase matching can be modeled by using a nonlinearity modulated with spatial period $\Lambda = 2\pi/K$,

$$\chi(z) = \chi_0 2\cos(Kz) = \chi_0 e^{-iKz} + cc, \tag{61}$$

neglecting high-order terms in the Fourier expansion, which are nonresonant. Denoting a gain constant as $\gamma = \omega_0 \chi_0 E_{p0}/c$, and dropping the nonresonant second term in Eq.(61), we have

$$\left[\frac{\partial}{\partial z} - ik(\omega)\right]\hat{a}(z,\omega) = i\gamma e^{i(k_p-K)z}\hat{a}^\dagger(z,2\omega_0-\omega). \tag{62}$$

The solution to Eq.(62) is

$$\hat{a}(z,\omega) = e^{i(k_p - K + k(\omega) - k(2\omega_0 - \omega))z/2} \left( f(\omega)\hat{a}(0,\omega) + g(\omega)\hat{a}^\dagger(0, 2\omega_0 - \omega) \right)$$
$$\approx e^{i(k_0 + k'(\omega - \omega_0))z} \left( f(\omega)\hat{a}(0,\omega) + g(\omega)\hat{a}^\dagger(0, 2\omega_0 - \omega) \right), \quad (63)$$

where $\Delta k(\omega) = k_p - K - k(\omega) - k(2\omega_0 - \omega)$ and $f$, $g$ are the same as in Eq.(8). In writing the input-output relation Eq.(7) we dropped the factor $\exp[ik_0 z + ik'vz]$, as it corresponds only to a common group delay during propagation through the nonlinear medium. That is, we denote $\hat{a}(z,\omega) = \exp[ik_0 z + ik'(\omega - \omega_0)z]\hat{b}(\omega)$ and $\hat{a}(\omega) = \hat{a}(0,\omega)$.

**Appendix B: ETPA with distinguishable photons**

For Type-II phase matching or noncollinear Type-0 or Type-1 PDC, the photons are distinguishable by virtue of their polarization, their direction of propagation, or both. In these cases the derivation in [32] can be generalized easily, as sketched here. See also [45]. Photons (or modes) may be classified as signal (*s*) or idler (*i*), for which there are separate creation operators, which commute between types. The two-mode squeezing transformation is generalized to broadband fields as

$$\hat{b}_s(\omega) = f(\omega)\hat{a}_s(\omega) + g(\omega)\hat{a}_i^\dagger(2\omega_0 - \omega)$$
$$\hat{b}_i(\omega) = f(\omega)\hat{a}_i(\omega) + g(\omega)\hat{a}_s^\dagger(2\omega_0 - \omega), \quad (64)$$

where $[\hat{a}_j(\omega), \hat{a}_k^\dagger(\omega')] = 2\pi\delta(\omega - \omega')\delta_{jk}$. The $f$ and $g$ functions are the same as in Eq.(8) for noncollinear Type-0 or Type-1 PDC, whereas for Type-II they need to be modified to include first-order dispersion (group-velocity mismatch), which typically results in a narrower spectrum. The time-gated operators can be shown to be

$$\hat{c}_s(\omega) = f(\omega)\hat{A}_s(\omega) + g(\omega)\hat{A}_i^\dagger(2\omega_0 - \omega)$$
$$\hat{c}_i(\omega) = f(\omega)\hat{A}_i(\omega) + g(\omega)\hat{A}_s^\dagger(2\omega_0 - \omega), \quad (65)$$

where

$$\hat{A}_s(\omega) = \int d\omega' W(\omega - \omega')\hat{a}_s(\omega')$$
$$\hat{A}_i(\omega) = \int d\omega' W(\omega - \omega')\hat{a}_i(\omega'), \quad (66)$$

with $[\hat{A}_s(\omega), \hat{A}_i^\dagger(\tilde{\omega})] = 0$ and $[\hat{A}_s(\omega), \hat{A}_s^\dagger(\tilde{\omega})] = [\hat{A}_i(\omega), \hat{A}_i^\dagger(\tilde{\omega})] = T \, \text{sinc}[(\omega - \tilde{\omega})T/2]$. The number of photons in each field is $N_s = N_i = T\int d\omega |g(\omega)|^2$.

The four-frequency correlation function is again given by Eq.(5), in which now $\hat{c}^\dagger(\omega) = \hat{c}_s^\dagger(\omega) + \hat{c}_i^\dagger(\omega)$. Of the four terms in in the correlation function, the ones that contain frequency anticorrelations and thus enhanced TPA are of the form

$$C_{si}^{(4)}(\omega',\tilde{\omega}',\omega,\tilde{\omega}) = \langle vac | \hat{c}_s^\dagger(\omega')\hat{c}_i^\dagger(\tilde{\omega}')\hat{c}_i(\omega)\hat{c}_s(\tilde{\omega}) | vac \rangle ,$$
$$= C_{si,coh} + C_{si,incoh}$$
(67)

where

$$C_{si,coh} \approx f^*(2\omega_0 - \omega')g^*(\omega')f(\omega)g(2\omega_0 - \omega)D(\tilde{\omega}' + \omega' - 2\omega_0)D(\omega + \tilde{\omega} - 2\omega_0)$$
$$C_{si,incoh} = g^*(\tilde{\omega}')g^*(\omega')g(\omega)g(\tilde{\omega}) \times$$
$$\langle vac | \hat{A}_i(2\omega_0 - \tilde{\omega}')\hat{A}_i^\dagger(2\omega_0 - \tilde{\omega}) \cdot \hat{A}_s(2\omega_0 - \omega')\hat{A}_s^\dagger(2\omega_0 - \omega) | vac \rangle$$
$$= g^*(\tilde{\omega}')g^*(\omega')g(\omega)g(\tilde{\omega})D(\tilde{\omega} - \tilde{\omega}')D(\omega' - \omega)$$
(68)

Note there is no added term of the form $D(\omega - \tilde{\omega}')D(\omega' - \tilde{\omega})$ as there is in the collinear Type-0 or Type-I cases because the $\hat{A}_s(\omega), \hat{A}_i^\dagger(\tilde{\omega})$ operators commute. Thus the 'flag' appearing in Eq.(38) has value $\xi = 0$ in this case.

The other terms that contribute to TPA are of the form $\langle vac | \hat{c}_s^\dagger(\omega')\hat{c}_s^\dagger(\tilde{\omega}')\hat{c}_s(\omega)\hat{c}_s(\tilde{\omega}) | vac \rangle$ and $\langle vac | \hat{c}_i^\dagger(\omega')\hat{c}_i^\dagger(\tilde{\omega}')\hat{c}_i(\omega)\hat{c}_i(\tilde{\omega}) | vac \rangle$, which correspond to TPA by pairs of signal-only or idler-only photons. Because these combinations lack the benefit of frequency anticorrelation, their contributions are small, of the same order as would appear in TPA by broad-band thermal-like light. (See **Appendix C**.) This statement is consistent with the fact that the signal (or idler) field alone has thermal-like statistics. [46]

For example, one of these terms is, using the fact that the $\hat{A}_s(\omega), \hat{A}_i^\dagger(\tilde{\omega})$ operators commute,

$$C_{ss}^{(4)} = \langle vac | \hat{c}_s^\dagger(\omega')\hat{c}_s^\dagger(\tilde{\omega}')\hat{c}_s(\omega)\hat{c}_s(\tilde{\omega}) | vac \rangle$$
$$C_{ss}^{(4)} = g^*(\tilde{\omega}')g^*(\omega')g(\omega)g(\tilde{\omega}) \times$$
$$\langle vac | \hat{A}_i(2\omega_0 - \tilde{\omega}')\hat{A}_i(2\omega_0 - \omega')\hat{A}_i^\dagger(2\omega_0 - \omega)\hat{A}_i^\dagger(2\omega_0 - \tilde{\omega}) | vac \rangle ,$$
$$\approx |g(\omega)|^2 |g(\tilde{\omega})|^2 \left( D(\omega - \omega')D(\tilde{\omega}' - \tilde{\omega}) + D(\omega - \tilde{\omega}')D(\omega' - \tilde{\omega}) \right)$$
(69)

which is of the form of the incoherent term $C_{incoh}$ in Eq.(38), and so contributes little to the TPA rate.

**Appendix: C TPA with broadband thermal-like state**

To verify the statement in **Appendix B** that TPA by signal-only or idler-only photons is equivalent to TPA by broad-band thermal-like light, we represent the thermal light using a classical random process, which is known to be consistent with quantum theory because such light has a representation as a Glauber-Sudarshan P distribution that is positive and well behaved (it is simply Gaussian). [47]

Representing the random process as $E_{BB}(\omega)$ in the frequency domain, its two-frequency correlation function is delta-correlated because the process is stationary in time,

$$\langle E_{BB}^*(\omega')E_{BB}(\omega'')\rangle = 2\pi P(\omega')\delta(\omega'-\omega'') ,  \quad (70)$$

where $P(\omega)$ is the field's power spectrum. The time-gated field is $E(\omega) = \int d\omega' E_{BB}(\omega')W(\omega-\omega')$ and its two-frequency correlation function is

$$\begin{aligned} C^{(2)} &= \langle E^*(\omega_1)E(\omega_2)\rangle \\ &= \int d\omega' P(\omega')W(\omega_1-\omega')W(\omega_2-\omega') \\ &\approx P(\frac{\omega_1+\omega_2}{2})\int d\omega' W(\omega_1-\omega')W(\omega_2-\omega') \\ &\approx P(\frac{\omega_1+\omega_2}{2})\times W(\omega_1-\omega_2) \end{aligned} \quad (71)$$

where we used the fact that the spectrum is slowly varying and much broader than the gate function for a time gate duration much longer than the correlation time (inverse of spectral width) of the thermal light.

The two-frequency correlation function is, using the Gaussian-moment theorem,

$$\begin{aligned} C^{(4)} &= \langle E^*(\omega')E^*(\tilde{\omega}')E(\omega)E(\tilde{\omega})\rangle \\ &= \langle E^*(\omega')E(\omega)\rangle\langle E^*(\tilde{\omega}')E(\tilde{\omega})\rangle + \langle E^*(\omega')E(\tilde{\omega})\rangle\langle E^*(\tilde{\omega}')E(\omega)\rangle . \\ &\approx P(\omega)P(\tilde{\omega})\big(W(\omega'-\omega)W(\tilde{\omega}'-\tilde{\omega})+W(\omega'-\tilde{\omega})W(\tilde{\omega}'-\omega)\big) \end{aligned} \quad (72)$$

This result is of the same form as the incoherent term Eq.(69) or Eq.(38), thus verifying the claimed equivalence.

**Appendix D: Second-order intensity autocorrelation function $g^{(2)}(0)$**

The second-order intensity autocorrelation function for the instantaneous intensity at zero time delay $g^{(2)}(0)$ for broadband squeezed light is found from the four-frequency correlation function using the Fourier relation $\hat{E}^{(+)}(0) = L_0 \int \dbar\omega \, \hat{c}(\omega)$, giving, after some algebra,

$$
\begin{aligned}
g^{(2)}(0) &= \frac{\langle \hat{E}^{(-)}(0)\hat{E}^{(-)}(0)\hat{E}^{(+)}(0)\hat{E}^{(+)}(0) \rangle}{\langle \hat{E}^{(-)}(0)\hat{E}^{(+)}(0) \rangle^2} \\
&= \frac{\int \dbar\omega' \int \dbar\tilde{\omega}' \int \dbar\omega \int \dbar\tilde{\omega} \, \langle vac | \hat{c}^\dagger(\omega')\hat{c}^\dagger(\tilde{\omega}')\hat{c}(\omega)\hat{c}(\tilde{\omega}) | vac \rangle}{\left( \int \dbar\omega \int \dbar\tilde{\omega} \, \langle vac | \hat{c}^\dagger(\omega)\hat{c}(\tilde{\omega}) | vac \rangle \right)^2} \\
&= (1+\xi) + \frac{T^2}{N^2} \left| \int \dbar\omega \, f(\omega)g(\omega) \right|^2 \\
&= (1+\xi) + \frac{\left| \int \dbar\omega \, f(\omega)g(\omega) \right|^2}{\left( \int \dbar\omega \, |g(\omega)|^2 \right)^2}
\end{aligned}
\qquad (73)
$$

We note that the same result is obtained when using the field operators $\hat{b}(\omega)$ before the time gate because the gate duration is assumed large compared to the field's coherence time.

Group-velocity delay occurring in the PDC crystal can reduce $g^{(2)}(0)$ by a small amount, creating apparent differences with the standard formula for idealized single-mode squeezing for collinear Type-0 or 1 PDC, where photons are indistinguishable ($\xi = 1$), [22, 23]

$$
g^{(2)}(0)_{ideal} = 3 + \frac{1}{\bar{n}} \ . \qquad (74)
$$

We thus introduce a dispersive term to give the dispersion-compensated form,

$$
g^{(2)}(0)_{comp} = (1+\xi) + \frac{\left| \int \dbar\omega \, f(\omega)g(\omega) \exp[iD_2(\omega-\omega_0)^2] \right|^2}{\left( \int \dbar\omega \, |g(\omega)|^2 \right)^2} , \qquad (75)
$$

as stated in Eq.(51), where $D_2$ represents the second-order (group delay) of a dispersion-compensating device such as a prism pair.

In **Fig. 4** in the main text we plotted the compensated and uncompensated forms of $g^{(2)}(0)$ versus the mean total photon number per pulse. We found there that the compensated version does not drop below 3, and follows the expected idealized form $3 + 1/\bar{n}$ when we take the number of photons per mode to be approximated by

$$\bar{n} \approx \frac{N}{(3wT/4\pi)} = \frac{F}{(3w/4\pi)}, \tag{76}$$

consistent with Eq.(32) within a factor of $2\pi$.

This result confirms that the effective number of modes is well approximated by $M_{modes} = N/\bar{n} = (3w/4\pi)T \approx BT$. Equation (76) is not an empirical fit but is derived as follows. Using from Eq.(44)

$$\left|\int d\omega f(\omega)g(\omega)\right|^2 \approx \left(\frac{N}{T}\right)\frac{3}{4}\frac{w}{\pi} + \left(\frac{N}{T}\right)^2 \tag{77}$$

and from Eq.(32)

$$\bar{n} \approx \frac{N}{BT} = \frac{\int d\omega |g(\omega)|^2}{B} \tag{78}$$

we obtain, from Eq.(73),

$$g^{(2)}(0) = (1+\xi) + \left(\frac{T}{N}\right)^2 \left[\left(\frac{N}{T}\right)\frac{3}{4}\frac{w}{\pi} + \left(\frac{N}{T}\right)^2\right]$$

$$= (2+\xi) + \left(\frac{3}{4\pi}\frac{wT}{N}\right) = (2+\xi) + \left(\frac{BT}{N}\right). \tag{79}$$

$$= (2+\xi) + \left(\frac{1}{\bar{n}}\right)$$

This derivation is the first to our knowledge that verifies the idealized $g^{(2)}(0)$ result for broadband spectrally multi-mode squeezing using a realistic model and finding an excellent approximation to an otherwise complicated numerical evaluation of the dispersion-compensated case.

As we noted earlier, the TPA probabilities plotted in **Fig. 4** can also be optimized by including dispersion compensation, although the uncompensated results as plotted using the simpler closed forms Eq. (42) are already within 15% of the compensated results. In **Fig. 9** we show the probability with optimized dispersion compensation, where the discrepancy between numerical and our approximate form is reduced to less than 3% when assuming an optimum amount of dispersion compensation subsequent to the PDC crystal.

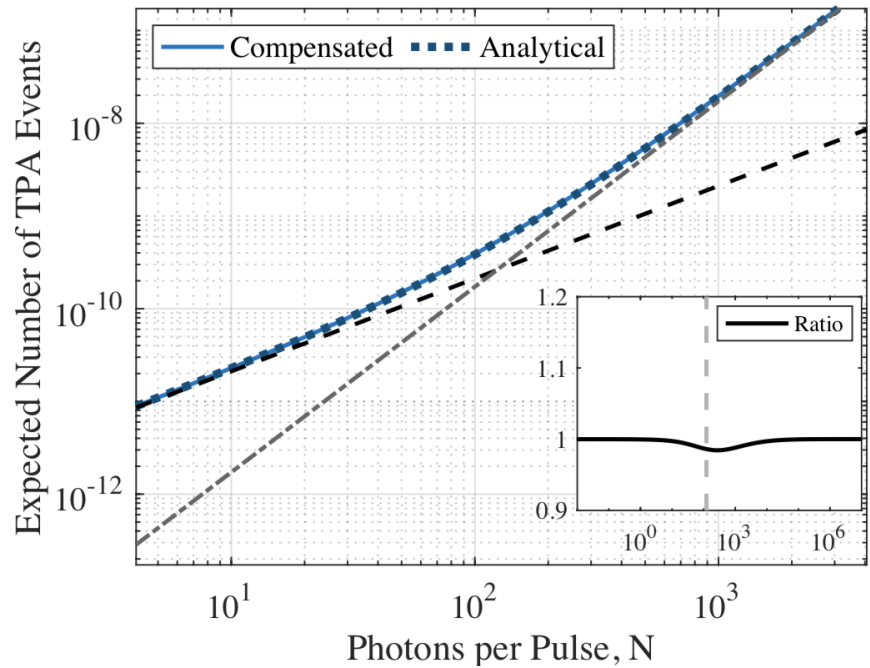

Fig. 9 Same as **Fig. 5** but with dispersion compensation implemented and optimized at every value of *N*, showing that the approximate analytical formula is within 3% of the numerical optimized result for all values of photon number.

**References**

[1] Landes, Tiemo, Michael G. Raymer, Markus Allgaier, Sofiane Merkouche, Brian J. Smith, and Andrew H. Marcus. "Quantifying the enhancement of two-photon absorption due to spectral-temporal entanglement." *Optics Express* 29, no. 13 (2021): 20022-20033.

[2] K. M. Parzuchowski, A. Mikhaylov, M. D. Mazurek, R. N. Wilson, D. J. Lum, T. Gerrits, C. H. Camp Jr, Martin J. Stevens, and R. Jimenez. "Setting Bounds on Entangled Two-Photon Absorption Cross Sections in Common Fluorophores." *Phys. Rev. Applied.* **15**(4), 044012 (2021).

[3] Landes, Tiemo, Markus Allgaier, Sofiane Merkouche, Brian J. Smith, Andrew H. Marcus, and Michael G. Raymer. "Experimental feasibility of molecular two-photon absorption with isolated time-frequency-entangled photon pairs." *Physical Review Research* 3, no. 3 (2021): 033154.

[4] Julio Gea-Banacloche, "Two-photon absorption of nonclassical light," *Physical review letters* **62**, 1603 (1989).

[5] Juha Javanainen and Phillip L. Gould. "Linear intensity dependence of a two-photon transition rate." *Physical Review A* **41**, 5088 (1990).

arXiv v 3; to appear Phys Rev A 2022                                                          July 8 2022

[6] H.-B. Fei, B. M. Jost, S. Popescu, B. EA Saleh, and M. C. Teich. "Entanglement-induced two-photon transparency." *Physical review letters* **78**(9), 1679-1682 (1997).

[7] Schlawin, Frank, and Shaul Mukamel. "Photon statistics of intense entangled photon pulses."*Journal of Physics B: Atomic, Molecular and Optical Physics* 46, no. 17 (2013): 175502.

[8] F. Schlawin, K. E. Dorfman, and S. Mukamel. "Entangled two-photon absorption spectroscopy." *Accounts of chemical research* **51**(9), 2207-2214 (2018).

[9] H. Oka, "Two-photon absorption by spectrally shaped entangled photons." *Physical Review A* **97**(3), 033814 (2018).

[10] Göppert-Mayer, "About elementary acts with two quantum leaps."*Annalen der Physik* **401**, no.3 (1931): 273-294

[11] Robert W. Boyd, *Nonlinear Optics*, 3rd Edition, Academic Press (Burlington, 2008), pg. 558.

[12] Raymer, Michael G., Tiemo Landes, Markus Allgaier, Sofiane Merkouche, Brian J. Smith, and Andrew H. Marcus. "How large is the quantum enhancement of two-photon absorption by time-frequency entanglement of photon pairs?" *Optica* 8, no. 5 (2021): 757-758.

[13] D.-I. Lee and T. Goodson. "Entangled photon absorption in an organic porphyrin dendrimer." *The Journal of Physical Chemistry B* **110**(51), 25582-25585 (2006).

[14] Tabakaev, Dmitry, Matteo Montagnese, Geraldine Haack, Luigi Bonacina, J-P. Wolf, Hugo Zbinden, and R. T. Thew. "Energy-time-entangled two-photon molecular absorption." *Physical Review A* 103, no. 3 (2021): 033701.

[15] Raymer, Michael G., Tiemo Landes, and Andrew H. Marcus. "Entangled Two-Photon Absorption by Atoms and Molecules: A Quantum Optics Tutorial." *J. Chem. Phys.* **155**, *081501 (2021); https://doi.org/10.1063/5.0049338*

[16] André Stefanov, "Harnessing entanglement from broadband energy entangled photon pairs," in Mukamel, Shaul, et al. "Roadmap on quantum light spectroscopy." *Journal of Physics B: Atomic, Molecular and Optical Physics* 53, no. 7 (2020): 072002.

[17] Cutipa, Paula, and Maria V. Chekhova. "Bright squeezed vacuum for two-photon spectroscopy: simultaneously high resolution in time and frequency, space and wavevector." *Optics Letters* 47, no. 3 (2022): 465-468.

[18] Barak Dayan, "Theory of two-photon interactions with broadband down-converted light and entangled photons." *Phys. Rev. A*, **76**: 043813, Oct 2007

[19] Drago, Christian, and John Sipe. "Aspects of Two-photon Absorption of Squeezed Light: the CW limit." *arXiv preprint arXiv:2205.07485* (2022).

[20] Spasibko, Kirill Yu, Denis A. Kopylov, Victor L. Krutyanskiy, Tatiana V. Murzina, Gerd Leuchs, and Maria V. Chekhova. "Multiphoton effects enhanced due to ultrafast photon-number fluctuations." *Physical review letters* 119, no. 22 (2017): 223603.

[21] Popov, A. M., and O. V. Tikhonova. "The ionization of atoms in an intense nonclassical electromagnetic field." *Journal of Experimental and Theoretical Physics* 95, no. 5 (2002): 844-850.

[22] Iskhakov, T. Sh, A. M. Pérez, K. Yu Spasibko, M. V. Chekhova, and G. Leuchs. "Superbunched bright squeezed vacuum state." *Optics Letters* 37, no. 11 (2012): 1919-1921.

[23] Loudon, Rodney, and Peter L. Knight. "Squeezed light." *Journal of modern optics* 34, no. 6-7 (1987): 709-759. Eq. 3.22.

[24] Scully M O and Zubairy M S 1997, *Quantum Optics* (Cambridge: Cambridge University Press)

[25] R. Mollow, "Two-Photon Absorption and Field Correlation Functions," Phys. Rev. 175, 1555 (1968)

[26] F. Schlawin, "Entangled photon spectroscopy." *Journal of Physics B: Atomic, Molecular and Optical Physics* **50**(20), 203001 (2017).

[27] Keller, Timothy E., and Morton H. Rubin. "Theory of two-photon entanglement for spontaneous parametric down-conversion driven by a narrow pump pulse." *Physical Review A* 56, no. 2 (1997): 1534.

[28] So-Young Baek and Yoon-Ho Kim. "Spectral properties of entangled photon pairs generated via frequency-degenerate type-I spontaneous parametric down-conversion." *Physical Review A* **77**, no. 4 (2008): 043807.

[29] Wasilewski, Wojciech, Alexander I. Lvovsky, Konrad Banaszek, and Czesław Radzewicz. "Pulsed squeezed light: Simultaneous squeezing of multiple modes." *Physical Review A* 73, no. 6 (2006): 063819.

[30] Christ, Andreas, Benjamin Brecht, Wolfgang Mauerer, and Christine Silberhorn. "Theory of quantum frequency conversion and type-II parametric down-conversion in the high-gain regime." *New Journal of Physics* 15, no. 5 (2013): 053038.

[31] Boitier, Fabien, Antoine Godard, Nicolas Dubreuil, Philippe Delaye, Claude Fabre, and Emmanuel Rosencher. "Two-photon-counting interferometry." *Physical Review A* 87, no. 1 (2013): 013844.

[32] Mostowski, J. and M. G. Raymer. "Quantum statistics in nonlinear optics," In *Contemporary nonlinear optics*, p. 187. Academic Press New York, 1992.

[33] Yuen, Horace P. "Two-photon coherent states of the radiation field." *Physical Review A* 13, 2226 (1976).

[34] Dayan, Barak, Avi Pe'er, Asher A. Friesem, and Yaron Silberberg. "Coherent control with broadband squeezed vacuum." *arXiv preprint quant-ph/0302038* (2003).

[35] Kopylov, Denis A., Kirill Yu Spasibko, Tatiana V. Murzina, and Maria V. Chekhova. "Study of broadband multimode light via non-phase-matched sum frequency generation." *New Journal of Physics* 21, no. 3 (2019): 033024.

[36] Dayan, Barak, Avi Pe'Er, Asher A. Friesem, and Yaron Silberberg. "Two photon absorption and coherent control with broadband down-converted light." *Physical review letters* 93, no. 2 (2004): 023005.

[37] Weiner, Andrew. *Ultrafast optics*. Vol. 72. John Wiley & Sons, 2011.

[38] Li, Z. W., C. Radzewicz, and Michael G. Raymer. "Phase cross correlation in the coherent Raman process." *Optics letters* 13, no. 6 (1988): 491-493.

[39] Li, Z. W., C. Radzewicz, and M. G. Raymer. "Cancellation of laser phase fluctuations in Stokes and anti-Stokes generation." *JOSA B* 5, no. 11 (1988): 2340-2347.

[40] Svidzinsky, Anatoly, Girish Agarwal, Anton Classen, Alexei V. Sokolov, Aleksei Zheltikov, M. Suhail Zubairy, and Marlan O. Scully. "Enhancing stimulated Raman excitation and two-photon absorption using entangled states of light." *Physical Review Research* 3, no. 4 (2021): 043029.

[41] Huttner, B., S. Serulnik, and Y. Ben-Aryeh. "Quantum analysis of light propagation in a parametric amplifier." *Physical Review A* 42, no. 9 (1990): 5594.

[42] Raymer, Michael G. "Quantum theory of light in a dispersive structured linear dielectric: a macroscopic Hamiltonian tutorial treatment." *Journal of Modern Optics* 67, no. 3 (2020): 196-212.

[43] Blow, K. J., Rodney Loudon, Simon JD Phoenix, and T. J. Shepherd. "Continuum fields in quantum optics. "*Physical Review A* 42, no. 7 (1990): 4102.

[44] R. Loudon, *The Quantum Theory of Light*, Oxford Science Publications, 3rd ed., Oxford University Press, Oxford 2000.

[45] Raymer, M. G., Jaewoo Noh, K. Banaszek, and I. A. Walmsley. "Pure-state single-photon wave-packet generation by parametric down-conversion in a distributed microcavity." *Physical Review A* 72, no. 2 (2005): 023825.

[46] Ou, Zheyu Jeff. *Quantum Optics for Experimentalists*. World Scientific (2017).

[47] Mandel, Leonard, and Emil Wolf. *Optical coherence and quantum optics*. Cambridge university press (1995).